\documentclass[twocolumn,pre,showpacs]{revtex4}
\usepackage{graphicx}
\usepackage{dcolumn}
\usepackage{amsmath}
\usepackage{latexsym}

\def\bea{\begin{eqnarray}}
\def\eea{\end{eqnarray}}
\def\beq{\begin{equation}}
\def\eeq{\end{equation}}
\def\f{\frac}

\def\ve{\varepsilon}

\def\be{\beta}
\def\D{\Delta}

\def\r{\rho}
\def\a{\alpha}
\def\s{\sigma}
\def\kb{k_B}
\def\la{\langle}
\def\ra{\rangle}
\def\nn{\nonumber}

\def\bv{{\bf v}}

\def\br{{\bf r}}

\def\d{\delta}
\def\p{\partial}
\def\l{\lambda}

\def\g{\gamma}

\def\a{\alpha}
\def\d{\delta}
\def\p{\partial} 
\def\nn{\nonumber}
\def\r{\rho}

\def\la{\langle}
\def\ra{\rangle}

\def\g{\gamma}

\def\f{\frac}

\begin{document}
\title{Novel Fluctuations at a Constrained Liquid-Solid Interface
}

\author{Abhishek Chaudhuri}
\affiliation{
Raman Research Institute, C. V. Raman Avenue, Sadashivanagar,
Bangalore - 560080, India
}
\email{abhishek@rri.res.in}
\author{Debasish Chaudhuri}
\affiliation{
Max Planck Institute for the Physics of Complex Systems, 
N{\"o}thnitzer Str. 38, 01187 Dresden, Germany
}
\email{debc@pks.mpg.de}
\author{Surajit Sengupta}
\affiliation{
S. N. Bose National Center for Basic Sciences, JD Block, Sector 3,
Salt Lake, Kolkata - 700098, India
}
\email{surajit@bose.res.in}
\date{\today}

\begin{abstract}
We study the interface between a solid 
trapped within a bath of liquid by a suitably shaped non-uniform external 
potential. Such a potential may be constructed using lasers, external 
electric or magnetic fields or a surface template. 
We study a two dimensional case where 
a thin strip of solid, created in this way, is surrounded on either side
by a bath of liquid with which it can easily exchange particles. 
Since height fluctuations of the interface cost energy, this interface 
is constrained to remain flat at all length scales. 
However, when such a solid is stressed by altering the depth of the potential; 
beyond a certain limit, it responds by relieving stress by novel 
interfacial fluctuations which involve addition or deletion of entire 
lattice layers of the crystal. This ``layering'' 
transition
is a generic feature of the 
system regardless
of the details of the interaction potential. We show how such 
interfacial fluctuations influence mass, momentum and energy transport 
across the interface. Tiny momentum impulses produce 
weak shock waves which 
travel through the interface and cause the spallation of 
crystal layers into the liquid. Kinetic and energetic constraints 
prevent spallation of partial layers from the crystal, a fact 
which may be of some practical use. We also study heat transport through
the liquid-solid interface and obtain the 
resistances in liquid, solid and interfacial regions (Kapitza resistance)
as the solid undergoes such layering transitions.
Heat conduction, which shows strong signatures of the structural 
transformations, can be understood using a 
free volume calculation.
\end{abstract}

%%Low-dimensional, mesoscopic, and nanoscale systems: structure and nonelectronic properties; Liquid-solid interfaces; High-pressure and shock wave effects in solids and liquids; Channel and internal heat flow
\maketitle

\section{Introduction}
Interfaces between co-existing phases in condensed matter systems posses an 
intrinsic width determined by competition between bulk and surface energies 
\cite{widom,vdwal,cahnhill,safran} which is usually further broadened by 
capillary fluctuations\cite{capil,binmul} viz. random 
fluctuations of the interface away from its mean position. The latter 
is a direct 
manifestation of the fact that translations of the interface in a direction
perpendicular to its plane cost no energy\cite{stanbar}. Such capillary 
fluctuations 
actually cause the interfacial width to diverge\cite{stanbar,edwil} with a 
cutoff determined 
by the system size. External potentials eg. gravitational fields or coupling
to a substrate however \cite{wern,capi-bin1,capi-bin2} tend to 
round-off this divergence and set additional length scales, absent in the 
free system. If one or both of the phases happens to be a solid, then 
long-ranged elastic energy costs for deforming the interface may also 
limit the broadening of the interface in equilibrium\cite{bin-supp}. Nevertheless, the analog 
of capillary fluctuations, viz. crystallization waves have been experimentally
observed at liquid-solid interfaces\cite{balibar}. 
It is, of course, also possible to suppress interfacial capillary 
fluctuations by position dependent chemical potentials which break the
translational symmetry of the interface explicitly\cite{ac-pas-ss}. One 
would expect that 
such interfaces would remain essentially flat over all length scales and 
therefore be completely inert. 

We show, in this paper, that this 
is not so. Surprisingly, there are novel
fluctuations and phenomena associated with such constrained interfaces
which have static as well as dynamic consequences.
Particles are transferred across the interface
in new and interesting ways. 
A liquid-solid interface constructed in such a fashion can have physically 
interesting height fluctuations while at the same time remaining flat! 
The system manages this by selecting only those fluctuations which 
involve changes in height by a single atomic spacing and have a wavelength 
equal to the size of system in the transverse direction. Further, such 
interfacial fluctuations are driven by the elastic response of the solid 
to stresses imposed by the external potential. As the depth of the trapping
potential is gradually increased, the solid accomodates this stress by 
incorporating layers of atoms from the liquid, either only parallel to the 
interface or alternately perpendicular and
parallel in a cyclical fashion, depending on the nature of the interactions. 
We believe that some of our predictions
may be directly checked for liquid-solid interfaces in atomic, as well
as, colloidal systems where the chemical potential field may be
provided either by a laser trap or by a patterned substrate. Needless to say, 
such fluctuations are expected to be observed only in systems where the overall 
size is small -- of the order of only a few atomic spacings in the transverse
direction. Preliminary results of our studies of this system have been 
published in Refs.\cite{nprl} and \cite{jpcm}.   

This paper is organized as follows. 
We begin, in the next section, by showing how 
a two-dimensional liquid-solid
interface may be produced using a non-uniform external potential. 
Monte Carlo (MC) computer simulations\cite{daan} in the constant
number, area and temperature (NAT) ensemble are set up to realize 
this explicitly 
for particles interacting with hard disk, soft disk and Lennard-Jones 
potentials. The interface is then characterized
using a variety of thermodynamic and structural
quantities, which are measured as a function of the perpendicular
distance from the interface. As a function of
the depth of the potential well, the trapped solid undergoes, what we
have called, ``layering" transitions\cite{aypa,collayer,nprl,jpcm,debc,debc-heat} which involve the addition (or
removal) of an entire layer of solid from (or into) the surrounding
liquid through the interface. This transition is described in detail for the 
hard disk system in section III. The layering transition is accompanied
by a sharp jump in the density of the solid. We obtain 
this density jump within a mean field, thermodynamic approach\cite{debc}.  
A comparison of the predictions of the thermodynamic 
theory and our MC computer experiments show the theory to be 
approximately correct to within a few percent. 
We show that the layering transition is a novel mechanism by which a stressed
nano-solid constrained by an external potential can respond
plastically to large stresses {\em without} nucleating
dislocations or cracks\cite{chasen,marder}. We establish that this 
phenomenon is general
and is independent of the particular interatomic potential used.
In section IV, we describe the kinetics of the layering transition.
In section V we explore how mass and
momentum are transported across the liquid-solid interface
and especially the role of the layering transition on the transport
coefficients. Molecular dynamics (MD) simulations in the constant number, 
area and energy (NAE) ensemble\cite{daan} are carried out for this purpose. 
We show in this paper that (1) fluctuations
associated with these transitions are of a special kind always
involving the transfer of complete layers of solid (2) these
fluctuations offer resistance to the transfer of momentum and energy
through the interface (3) the resistance is maximum when the energy
matches that required to raise a complete lattice layer from within
the potential well into the surrounding liquid. We study the stability 
of surface kinks at the liquid
solid interface in the hard disk system. We then study the
response of the interface to weak acoustic shocks\cite{zeldovich} which 
are shown to cause {\em spallation} of complete lattice layers if the incident 
energy of the shock is large enough. 
In section VI we use non-equilibrium molecular dynamics to study heat
transport\cite{debc-heat} through the liquid-solid interface in soft disks 
and obtain the conductance of the system in different regions, 
heat current and contact or Kapitza resistance\cite{barrat,kapitza} of the 
interface as a function of the depth of the potential well. 
The heat conduction is particularly sensitive to the fluctuations
in the direction of current flow. We present an approximate free volume
type calculation that qualitatively captures the response of the heat
conductance to the internal changes in the solid induced by the trapping
potential\cite{debc-heat}.  Finally, in section VII we conclude 
after discussing some consequences of our study and its relevance to
experiments.

\section{Constrained Liquid-Solid Interface: Static properties}
\label{hdint}
\begin{figure}[t]
\begin{center}
\includegraphics[width=7.0cm]{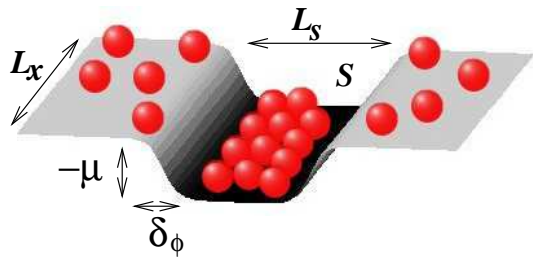}
\end{center}
\caption{(Color online)
A schematic diagram of the system showing the liquid and 
solid regions produced by the external chemical potential of depth $-\mu$.
The various dimensions mentioned in the text are also marked in the figure.}
\label{tanhp}
\end{figure}

In this section, we explore the possibility of creating a patterned sequence
of confined solid and liquid regions using an external, space-dependent,
chemical potential field $\phi({\bf r})$.
Consider a two dimensional system (see Fig.\ref{tanhp}) of $N$ atoms
of average density (packing fraction) $\eta = \pi N/4 A$ within
a rectangular cell of size $A = L_x \times L_y$ where the central region 
${\cal S}$ of area $A_s = L_x \times L_s$ is occupied by $N_s$
atoms arranged as a crystalline solid of density $\eta_s > \eta$, 
while the rest of the cell is filled with liquid of density
$\eta_\ell < \eta$. 
The difference in density is produced by an external 
field $\phi({\bf r}) = -\mu$ for ${\bf r} \in {\cal S}$; increasing sharply 
but smoothly 
to zero elsewhere with a hyperbolic tangent profile of width $\delta_{\phi}$. 
How may $\phi({\bf r})$ be realized in practice? In model solids like 
colloids \cite{hamley}, one may use a surface template to create a static 
pattern \cite{AMOLF}. In real systems, as well as colloids \cite{baumgartl}, 
one may be able to use laser traps \cite{phillips} or non~-uniform electric 
or magnetic fields. Usual laser traps for alkali metals or rare gas atoms are 
in the range of $10mK$ for which a power of about $100mW$ is 
required\cite{metcalf}. For colloidal systems, required laser powers are 
even lower\cite{col-trap}.
 
We first describe our results for a system where the atoms interact with a   
{\em hard disk} potential \cite{jaster} which is infinitely repulsive 
if the distance $r_{ij}$ between two atoms $i$ and $j$ is less than $\sigma$, 
the hard disk diameter and zero otherwise. We show later 
that qualitative results for more realistic potentials\cite{daan}, e.g.,
soft disk or Lennard-Jones are similar.
We have chosen $\delta_{\phi} = \sigma/4$, where $\sigma$ is the hard disk
diameter and sets the scale of length. 
The energy scale for this system is set by $k_BT$ where $k_B$ is the 
Boltzmann constant and $T$ the temperature. In our simulations we
set $\sigma = 1$ and $k_BT = 1$ unless otherwise stated. 

\begin{figure}[t]
\begin{center}
\includegraphics[width=7.0cm]{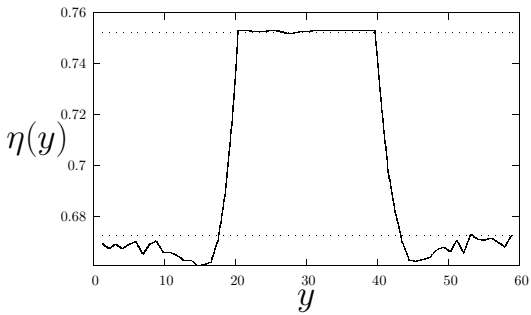}%{Figures/spalitza-fig2.eps}
\end{center}
\caption{The density
profile $\eta(y)$ coarse grained over strips of width $\sigma$
(averages taken over $10^3$ MC configurations each separated by $10^3$
MCS) at $\mu=6$. 
The dotted lines show the predictions for liquid density $ \eta_{\ell}=0.6725$
and solid density $\eta_s=0.752$ from a simple thermodynamic theory
presented in Sec.\ref{layer}.A.}
\label{dens}
\end{figure}

\begin{figure}[t]
\begin{center}
\includegraphics[angle=90,width=7.0cm]{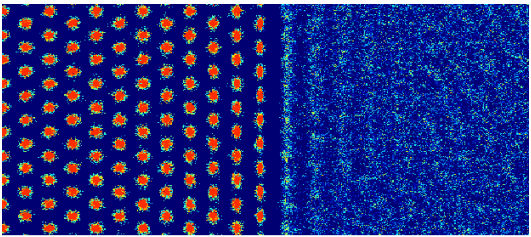}
\end{center}
\caption{(Color online)
Solid-liquid interface at $\mu=8$.
Superposition of 500 configurations separated by $10^3$ MCS showing a
solid like order (red : high $\eta$) gradually vanishing into the fluid
(blue : low $\eta$) across a well defined solid-fluid interface.}
\label{system}
\end{figure}

\begin{figure}[t]
\begin{center}
\includegraphics[width=7.0cm]{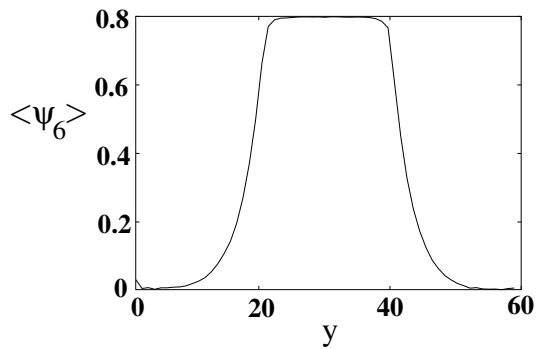}
\end{center}
\caption{Bond orientational order parameter across the liquid-solid
interface for a $21$ layered solid at $\mu=6$ surrounded by liquid on 
both sides.}
\label{border}
\end{figure}

\begin{figure}[t]
\begin{center}
\includegraphics[width=7.0cm]{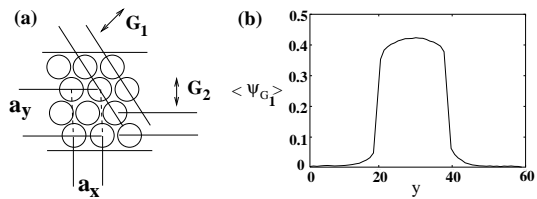}
\end{center}
\caption{(a)The reciprocal lattice vectors ${\bf G}_1$ and ${\bf G}_2$, and
the rectangular unit cell. (b) Solid order parameter corresponding to
${\bf G}_1$ is nonzero in the region ${\cal S}$ at $\mu=6$, indicating a solid. }
\label{g1g2}
\end{figure}

\begin{figure}[t]
\includegraphics[width=8.6cm]{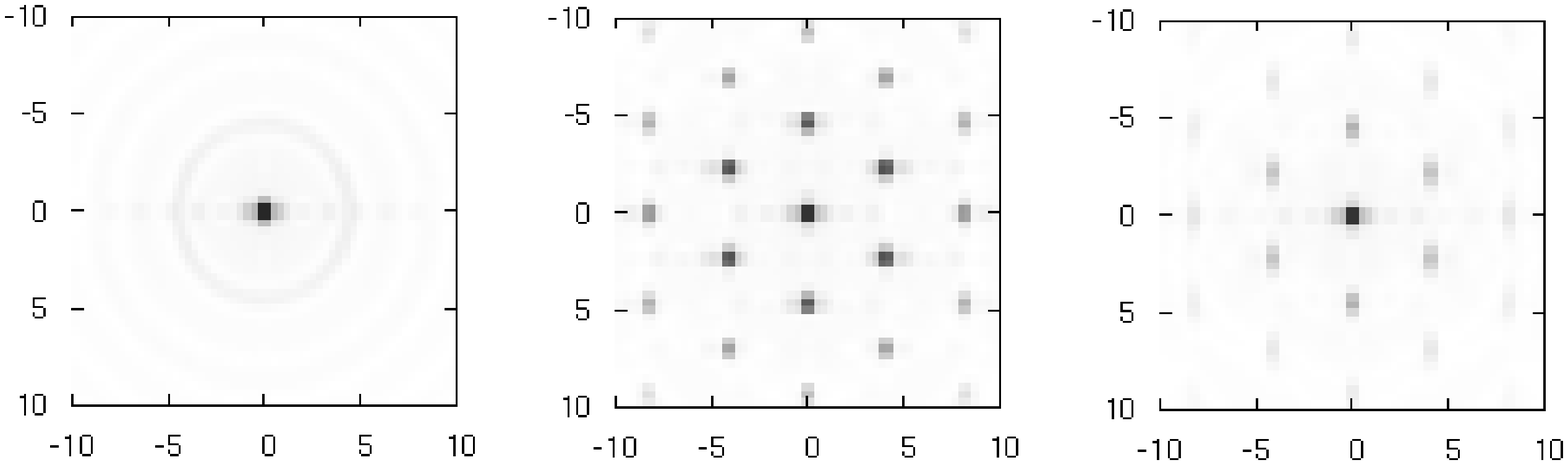}%{spalitza-fig6.eps}
\caption{From left to right, the three view-graphs show structure factors 
for the liquid, the solid and the interfacial regions respectively at $\mu=6$.} 
\label{sq}
\end{figure}

The full configuration dependent Hamiltonian is 
${\cal H} = \sum_{ij} u(\br_{ij}) + \sum_{i}\phi({\bf r}_i)$. We have carried 
out extensive MC simulations with usual Metropolis moves \cite{daan}, 
periodic boundary conditions in both directions and in the constant 
number, area and temperature ensemble to obtain the equilibrium behaviour of 
this system for different $\mu$ at a fixed average $\eta$. 
$N=1200$ particles occupy an area $A = 22.78 \times 59.18$ with the 
solid occupying the central third of the cell of size $L_s = 19.73$.
The initial configuration is chosen to be a liquid with 
$\eta = 0.699$; close to but slightly lower than the bulk
freezing density $\eta_f = 0.706$ \cite{jaster}. 
On equilibration, ${\cal S}$ contains a solid with the close-packed 
planes  parallel to the solid-liquid interfaces which lie, at all times,
along the lines where $\phi(y) \to 0$. The equilibration time 
is large and many ($\sim 10^7$) Monte Carlo steps (MCS) are discarded
before results shown in Figs.~\ref{dens} to \ref{step} are obtained.

The density $\eta(y)$, coarse grained over strips of width $\sim \sigma$, 
varies from its value $\eta_\ell$ in the liquid 
to $\eta_s$ as we move into the region ${\cal S}$ (Fig.\ref{dens}). 
Averages are taken over $10^3$ MC configurations each separated by $10^3$ MC steps.
The trap depth $\mu = 6$, supports an equilibrium solid of density 
$\eta_s =0.753$ in contact with a fluid of density $\eta_{\ell} =0.672$. 
The horizontal lines are predictions of a simple free-volume based theory 
(\cite{nprl}) for $\eta_s$ and $\eta_{\ell}$. We discuss the theory in 
Sec. III.A.  A superposition of atomic positions 
shows a static, flat, liquid-solid interface with the 
solid like order gradually vanishing into the liquid (Fig.~\ref{system}).
We have thus created a thin nano-sized crystal which is $21$ atomic
layers wide (for a trap depth, $\mu = 6$) and is flanked on either side by 
liquid separated by two liquid-solid interfaces.

The bond orientational order parameter $\langle \psi_6(y) \rangle$ 
where, the local value of $\psi_6$ for a particle $i$ 
located at ${\bf r}_i = (x,y)$ is given by
\begin{eqnarray}
\psi_{6,i} = \frac{1}{N_i}\sum_j \exp(6i\theta_{ij})\, .
\end{eqnarray}
The sum is over the $j \in N_i$ neighbours of $i$-th particle, and
$\theta_{ij}$ is the angle between the vector $\br_{ij}$ and an arbitrary
but fixed reference axis. To obtain $\langle \psi_6(y) \rangle$, this 
quantity is coarse grained over strips
of width $\sigma$ (averages taken over $10^3$ MC configurations each separated
by $10^3$ MCS). 
This shows a sharp rise from zero to a value close to one, as we
move into the region $\cal S$ (Fig.\ref{border}).
This indicates that the particles in $\cal S$ maintains hexatic order.
However, this does not necessarily justify the phase to be
a solid. Therefore, there is the need to calculate the solid order parameter. 

The order parameters corresponding to the solid phase
are the Fourier components of the (nonuniform) density-density correlation
$\langle \rho({\bf r}_i)\rho({\bf r}_j) \rangle$ calculated at the
reciprocal lattice points $\{{\bf G}\}$. This (infinite) set of numbers are all
zero (for ${\bf G} \ne 0$) in a uniform liquid phase and nonzero in a solid.
We restrict ourselves to the star consisting of the six smallest reciprocal
lattice vectors of the two dimensional triangular lattice. In modulated
liquid phase, the Fourier components corresponding to two out of these six
vectors, eg., those in the direction perpendicular to the interface,
${\bf G}_2$, are nonzero. The other four
components of this set which are equivalent by symmetry (${\bf G}_1$)
are zero in the (modulated) liquid and
nonzero in the solid (if there is true long ranged order). Thus, we use the
following order parameter :
\begin{eqnarray}
\langle \psi_{{\bf G}_k} \rangle = \f{1}{N^2}
\langle |\sum_{i,j=1}^N \exp(-i {\bf G}_k 
\cdot {\bf r}_{ij})|\rangle
\end{eqnarray}
where ${\bf r}_{ij} = {\bf r}_i - {\bf r}_j$. The solid order parameter 
so defined, in the direction ${\bf G}_1$ (see Fig.~\ref{g1g2}) and the
others equivalent to it by symmetry is nonzero in the region $\cal S$,
indicating the nucleation of a solid phase.
Note that $\langle \psi_6 \rangle$
shows a larger interfacial region than that obtained from
$\langle \psi_{{\bf G}_1} \rangle$. This is because the liquid near the liquid
solid interface is orientationally highly ordered due to the proximity of
the solid.

The structure factor describes density correlations in Fourier space,
\begin{eqnarray}
S(k) = N^{-1}\langle \rho(k)\rho(-k) \rangle
\end{eqnarray}
where $\rho(k) = \sum_{i=1}^{N}\exp(i{\bf k}\cdot{\bf r}_i)$. 
In a simulation with periodic boundaries, ${\bf k}$ is restricted by 
the periodicity of the system, i.e. with the simulation box.
The two-dimensional structure factor shows sharp peaks at triangular lattice 
positions for the solid region and isotropic ring pattern for the 
liquid (Fig.~\ref{sq}). 
The interfacial region also shows diffuse peaks at approximately triangular
lattice positions indicating once 
more that the interfacial region has considerable orientational and 
short ranged translational order.

Before we end this section, we  must voice a note of caution about 
the identity of phases in small and confined systems similar to the 
one we have here. It is well known, for example, that in two dimensions
it is impossible to obtain a solid with true long ranged order\cite{chaikin}.
Displacement correlations defined as 
$\langle {\bf u}(r)\,\cdot\,{\bf u}(0) \rangle$, 
where the displacement vector ${\bf u}$ is measured with respect to the 
zero temperature perfectly crystalline reference solid, grow as 
$\log(r)$.  
This implies that 
true Bragg peaks are impossible in two dimensions. Nevertheless, finite size 
and lack of complete averaging can lead to structure factors with sharp peaks,
while obtaining the true logarithmic divergence may need considerable
amount of computational effort. For a solid confined to a two-dimensional 
channel, the situation is even more dramatic. The displacement correlations 
now increase linearly with system size\cite{andrea} for distances larger 
than a crossover length $\sim L_s\,\log\,(L_s)$. The structure factor 
should show true Bragg peaks for reflections from planes parallel to the 
confining walls and should be diffuse in the other directions, implying 
therefore {\em smectic} like ordering throughout. In this paper, however, 
we continue to use the words ``solid'' and ``liquid'' in the usual sense 
referring to the presence or lack of solid-like order as shown in 
Fig.\ref{system} and Fig.\ref{g1g2}. This is mainly due to the fact that
our motivation of this study is to probe and understand the properties
of small (nano) systems. For such systems with channel length $L_x$ not
much larger than $L_s$ the phonon fluctuations are suppressed and can 
not destabilize the solid.
However, even for very large strips where phonon fluctuation destabilizes
a solid to a {\em smectic} phase, layering transitions 
are expected to occur\cite{degennes} and some associated proterties 
like accoustic spallation of layers and large change in heat conduction, 
which we describe in this paper, are expected to remain operative.
We must, at the same time, keep in mind
that some of the properties of the confined solid, including the layering 
transition and the ease of spallation of solid layers in response to 
weak acoustic shock (to be discussed below), may in-fact, be a consequence 
of incomplete ordering.    
 
\section{The layering transition}
\label{layer}
\begin{figure}[t]
\begin{center}
\includegraphics[width=7.0cm]{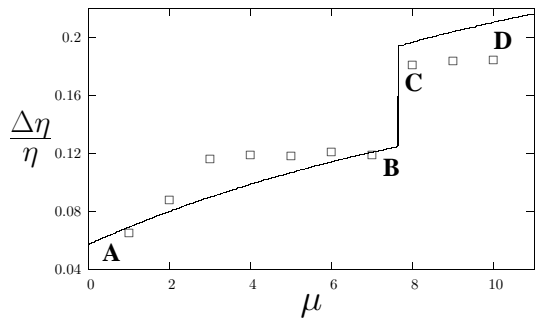}%{Figures/spalitza-fig7.eps}
\end{center}
\caption{ 
Plot of the equilibrium fractional density change $\Delta\eta/\eta$ as a
function of $\mu$ (points (MC data), solid line (free volume theory),
showing discontinuous jump at $\mu \approx 8$. The labels A-D mark the stable
21 layer solid (A), the transition (B-C) and the stable 22 layer solid (D)
respectively.}
\label{step}
\end{figure}

We now calculate the difference in densities between the solid and liquid
regions $\Delta \eta = (\eta_s - \eta_\ell)$ as a function of the strength of
the external field $\mu$. 
While $\Delta\eta/\eta$ increases with increasing $\mu$ as expected, the 
smooth increase is punctuated by a sharp jump (Fig.~\ref{step}). An examination
of the
\begin{figure}[t]
\begin{center}
\includegraphics[width=7.0cm]{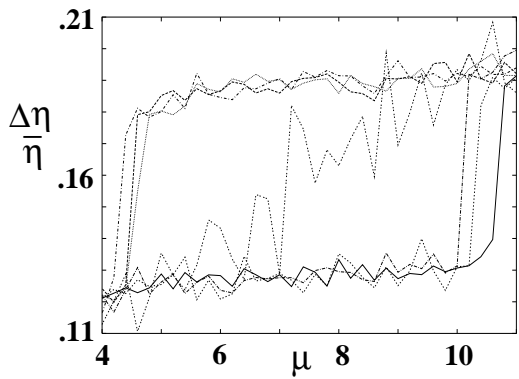}
\end{center}
\caption{Hysteresis loop as $\mu$ is cycled at the rate of $0.2$
per $10^6$ MCS. The central jagged line is the result of the initial cycle
when a single dislocation pair was present in the solid region.}
\label{hyst}
\end{figure} 
particle configuration shows that the jump occurs when an extra close~-packed 
layer enters ${\cal S}$ increasing the number of solid layers by one.
For the parameters in our simulation, the jump occurs at $\mu \approx 8$ with
the number of layers increasing from $21$ to $22$.
The value of $\mu$ at the jump is a strong function of $L_s$.
The solid structure is seen to be a defect free triangular lattice  
with a small rectangular distortion $\varepsilon_d(\eta_s,L_s)\,$\cite{debc}. 
We have examined the variation of $\Delta\eta(\mu)/\eta$ by cycling $\mu$ 
adiabatically around the region of the jump. This yields a prominent 
hysteresis loop as shown in Fig.~\ref{hyst} which indicates that 
`surface' steps (dislocation pairs) nucleated in the course of adding 
(or subtracting) a solid layer, have a vanishingly short lifetime.
Consistent with this we find that the jump in
$\Delta\eta(\mu)/\eta$ vanishes when the system is minimised at each $\mu$ 
with a constraint that the solid contains a single dislocation pair 
(Fig.~\ref{hyst}).
Interestingly, a dislocation pair forced initially into the bulk, rises 
to the solid-fluid interface due to a gain in strain energy \cite{lanlif1}, 
where they form surface indentations flanked by kink-antikink pairs. This costs
energy due to the confining potential, as a result, the kink-antikink pair gets
quickly annihilated by incorporating particles from the adjacent fluid.
The jump in  $\Delta\eta(\mu)/\eta$ is also seen to decrease with increasing 
$\delta_{\phi}$.

\subsection{The thermodynamic theory}
The qualitative features of these results may be obtained by a simple
thermodynamic theory (Fig.~\ref{step}) with harmonic distortions of the solid,
ignoring contributions from spatial variations of the density.
Note that this theory is in the spirit of a mean field approximation where 
all effects of fluctuations discussed in the concluding part of 
the last section are ignored. The free energy of the total system is 
written down as a sum over the free energies of the solid and the 
fluid. The free energies of the bulk hard disk fluid and solid are 
relatively easy to obtain as shown below.

\begin{figure}[t]
\begin{center}
\includegraphics[width=5.0cm]{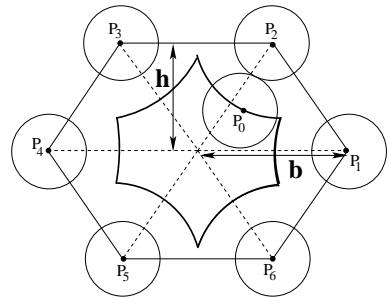}
\end{center}
\caption{In our free-volume calculation we assume that the outer six disks
  are fixed at their average positions
and the central disk moves within this cage of fixed particles. 
The curve in bold line shows the boundary $\cal{B}$ of the free
volume available to the central test particle. 
A point on this boundary is denoted by $P_0(x,y)$ while the
centers of the six fixed disks are denoted by $P_i(x_i,y_i)$ with
$i=1,2...6$. $b$ and $h$ denotes the base and the height
which uniquely decides the perimeter of $\cal {B}$ and the enclosed 
free volume. These are functions of density $\eta_s$ and width of the 
potential well $L_s$.
}
\label{fv}
\end{figure}

\subsubsection{Free energy of the solid}
In order that the solid channel accommodates
$n_l$ layers of a homogeneous triangular lattice with lattice parameter
$a_0$ of hard disks of diameter $\sigma$, the channel width
\begin{eqnarray}
L_s = \frac{\sqrt3}{2}(n_l - 1)a_0 + \sigma \, .
\label{els}
\end{eqnarray}
Defining
\begin{eqnarray}
\chi(\eta_s,L_s) = 1 + \frac{2(L_s - \sigma)}{\sqrt3 a_0} \, ,
\label{chi}
\end{eqnarray}
if $\chi = integer = n_l$ Eq.\ref{els} is recovered and the channel can
accomodated $n_l$ layers of homogeneous triangular lattice and
$\chi \neq integer$ implies a rectangular strain away from the perfect
trianglular lattice. 
For any given set of values for $L_s$ and $\eta_s$ one can find $\chi$ in
the following manner. One can associate a triangular lattice of 
lattice parameter $a_0$ from any given $\eta_s=\pi/2\sqrt 3 a_0^2$. This
set of $a_0$ and $L_s$ defines a specific $\chi$. Then the channel can
accommodate $n_l=int(\chi)$ (nearest integer to $\chi$)
number of layers of a centered rectangular (CR)
lattice with lattice parameters $a_y = 2(L_s - \sigma)/(n_l - 1)$ and,
$a_x = \pi/2 \eta_s a_y$. This lattice has strains 
$\ve_{xx}=\frac{n_l - 1}{\chi - 1}-1$ and
$\ve_{yy}=\frac{\chi - 1}{n_l - 1}-1$.
The deviatoric strain $\varepsilon_d = \varepsilon_{xx} -
\varepsilon_{yy}$ is then,
\begin{eqnarray}
\varepsilon_d = \frac{n_l - 1}{\chi - 1} - \frac{\chi - 1}{n_l - 1}\, .
\end{eqnarray}

In the fixed neighbour free volume theory (FNFVT), particles of high
density solid are assumed to be confined within a cage formed by the
average positions of its nearest neighbours. This cage and the available
free volume of the test particle to move around in this cage is entirely
defined by the quantities $b=a_0(1+\ve_{xx})$ and 
$h=(\sqrt 3 a_0/2)(1+\ve_{yy})$ (Fig.\ref{fv})\cite{debc-heat}. 
The amount of available
free volume $v_{fv}(\eta_s,L_s)$ bounded by the bold line $\cal B$ in
Fig.\ref{fv}, can be calculated by using tedious but 
straightforward geometrical considerations. The free volume 
free energy density is $f_s(\eta_s,L_s) = -(4\eta_s/\pi)\, \kb T \log (v_{fv})$.
$f_s$ always remains an upper bound to the exact free energy. This
upper bound becomes assymptotically exact in the close-packed limit.
Since $\ve_{xx}$ and $\ve_{yy}$ have discontinuity at half integral $\chi$,
the free energy $f_s$ has maxima at those $\chi$ values\cite{debc}.
In Sec.\ref{heat} we present an approximate theory for calculating heat
conductance in solid using this free volume approach\cite{debc-heat}.

\subsubsection{Free energy of the liquid}
The free energy density of the liquid bulk phase may be simply written as
\begin{eqnarray}
f_\ell = \r_\ell\int_0^{\eta_\ell} d\eta_\ell^{\prime}\frac{P/\rho_\ell - 1}{\eta_\ell^{\prime}} + f_{id}
\end{eqnarray}
where $\r_\ell=4\eta_\ell/\pi$, the ideal gas Helmholtz free energy density 
$f_{id} = \r_\ell\log(\rho_\ell) - \r_\ell$
and we use the semiemperical equation of state for the hard disk liquid
in Ref.\cite{santos}. This equation of state has been observed to show 
excellent agreement with hard disk fluid up to $\eta_\ell=0.65$ even when 
the fluid is confined in a hard narrow channel\cite{debc-prep}.
Following Ref.\cite{santos} we use $f_\ell = f_{san}+f_{id}$ where
\bea
f_{san} = \r_\ell
\frac{(2 \eta_c -1)\log\left(1-(2 \eta_c -1)\frac{\eta}{\eta_c}\right)
             -\log(1-\frac{\eta}{\eta_c})}{2(1-\eta_c)}\nn\\
\eea
with $\eta_c=\pi/2\sqrt 3$, the close-packed density for 2D hard disk 
triangular lattice.

\subsubsection{Free energy of the system}
We now write down the total free energy density of the system (fluid + solid
regions) using the free energy density expressions for solid and liquid bulk
phases as
\begin{eqnarray}
f = x \left[ f_s(\eta_s,L_s) - 4 \eta_s \mu/\pi\right] + 
(1-x) f_\ell(  \eta_{\ell}).
\label{}
\end{eqnarray}
We then minimise this free energy density with the constraint that the average
density is fixed,
$\eta = x \eta_s + (1-x)   \eta_{\ell}$, where $x$ is the area fraction
occupied by ${\cal S}$. The result of this calculation is
shown in Fig.~\ref{step} where it is seen to reproduce the jump
in $\Delta\eta(\mu)/\eta$.

\begin{figure}[t]
\begin{center}
\includegraphics[width=7.0cm]{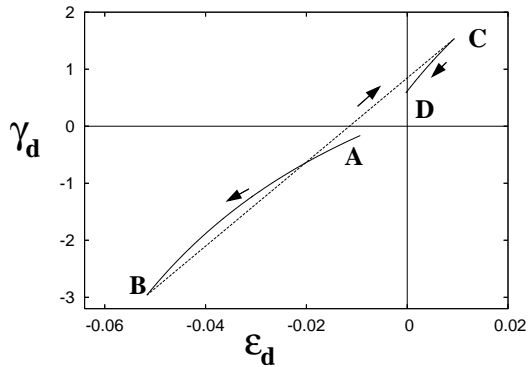}%{Figures/spalitza-fig9.eps}
\end{center}
\caption{
A plot of the deviatoric stress $\gamma_d$ against strain $\varepsilon_d$. The
arrows show the behaviour of these quantities as $\mu$ is increased from the
points marked A -- D. The labels correspond to the same states as in 
Fig.\ref{step}.
}
\label{stsn}
\end{figure}

Why does the solid incorporate layers of atoms from the liquid ? This
question may be answered elegantly if one calculates the deviatoric stress
$\g_d$ in the solid region as a function of the depth of the strain, 
$\varepsilon_d$. The stress may in fact be obtained in a straight forward 
fashion from the expression of the free energy. Differentiating the free 
energy of the solid with respect to $\varepsilon_d$ we obtain
\begin{eqnarray}
\gamma_d = \frac{\partial f_s}{\partial \varepsilon_d}.
\end{eqnarray}
When $\gamma_d$ is plotted versus the deviatoric strain $\varepsilon_d$, we
observe that the solid is not stress free for any arbitrary 
combination of $\mu$ and $L_s$. In fact, for our parameters, initially the $21$
layered solid is under tension in $y$- direction. 
We follow the variation of the deviatoric stress
with the strain as $\mu$ is increased from the points $A-D$ in the
Fig.~\ref{stsn}.
The state of stress in the solid jumps discontinuously from tensile to
compressive from B\,$\rightarrow$\,C due to
an increase in the number of solid layers by one accomplished by
incorporating particles from the fluid. This transition is reversible
and the system relaxes from a state of compression to tension by ejecting
this layer as $\mu$ is decreased.
As $\mu$ is increased, the tension increases
till $\g_d$ reaches about $-2.96$ when the corresponding strain is about 
$-0.052$.
At this point a layer enters the solid region and the stress and strain
switches from tensile to compressive. Further increase in $\mu$ now
decreases the stress and drives the solid to a state of zero stress at $\mu=10$.
Thus, the layering transition from $21$ to $22$ layers as observed by us is a
mechanism for relieving stress.

In our theory we assumed the channel can accomodate only an integer
number of layers and we ignored any possibility of defects. However, our
theory allowed for continuous change of base length $b$. In small solid,
this change is far from continuous. In our FNFVT, while $h$ 
remains constant until a layering transition, $b$ continuously reduces 
with increase in $\mu$. In reality, $b$ can only be reduced by
incorporating a new particle in an existing layer. Once a solid is formed
(for $\mu>4$, see Fig.\ref{step}), formation of point or line defects are
energetically very costly, thus even on an average $b$ can reduce only if
all the layers include a particle each and lattice parameters shrink in coherence. This is the
reason why in theoretical prediction $\D\eta/\eta$ grows smoothly in between
the layering transitions, though the same quantity remains almost constant
in simulation (Fig.\ref{step}). We shall see in later sections that the
coherent absorption of particles by all the layers at a time is indeed 
observed at larger well depth $\mu$ in similar simulations with soft core
particles.

\subsection{Layering in other potentials}
If the layering transition observed in the hard disk system is actually a new 
mechanism for relieving stress in a thin crystal, it should be independent
of the details of the potential. Our main results trivially extend to particles
interacting with any form of repulsive potential, or even when the interactions
are augmented by a short range attraction, provided we choose $\mu$ deeper than
the depth of the attractive potential.
\begin{figure}[t]
\begin{center}
\includegraphics[width=7.0cm]{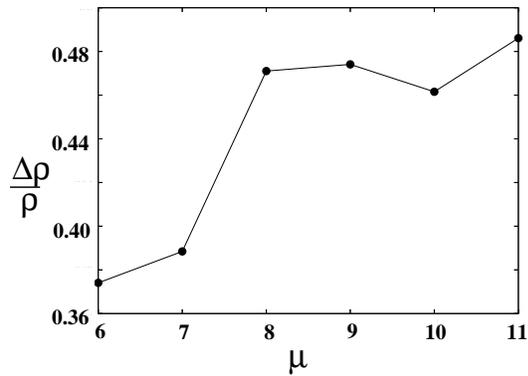}%{Figures/spalitza-fig10.eps}
\end{center}
\caption{ 
Plot of the equilibrium fractional density change $\Delta\rho/\rho$ as a 
function of $\mu$ (points -- MC data for soft core; line is a guide to eye).}
\label{step-soft}
\end{figure}

\begin{figure}[t]
\begin{center}
\includegraphics[width=7.0cm]{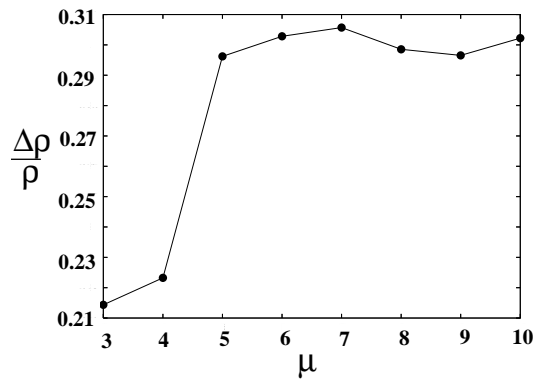}%{Figures/spalitza-fig11.eps}
\end{center}
\caption{ 
Plot of the equilibrium fractional density change $\Delta\rho/\rho$ as a 
function of $\mu$ (points -- MC data for Lennard-Jones system; line is 
a guide to eye).}
\label{step-lj}
\end{figure}
In this section we show explicitly that the layering transition is present
in the soft core and Lennard-Jones systems. We choose $\kb T=1$.
Once again we perform MC simulations
in the constant ${NAT}$ ensemble with periodic boundary conditions and
with the external chemical potential $\mu$. The relevant parameters 
corresponding to these potentials, namely $\epsilon$ and $\sigma$ in 
$u^{SS}(r) = \epsilon\left(\frac{\sigma}{r}\right)^{12} = ar^{-12}$ and
$u^{LJ}(r) = 4\epsilon\left[\left(\frac{\sigma}{r}\right)^{12} - 
\left(\frac{\sigma}{r}\right)^{6}\right]$, 
set the energy scale and the length scale respectively. Here $r = r_{ij}$ is
the distance between the pair of atoms $i,j$.
In our simulation $\epsilon = \sigma = 1$ and ${N} = 1200$ particles
occupy an area ${A} = 24\times 60$ with the solid occupying the central
third of the cell of size $L_s = 20$. The average density of the system is
therefore kept at $\rho = 0.833$ which is to be compared with the 
freezing densities  $\rho\simeq 0.987$~\cite{bgw} and 
$\rho\simeq 0.865$~\cite{tox} for the soft core and Lennard-Jones systems 
respectively. For the soft core potential, the $\mu$ value at the jump in 
density (Fig.~\ref{step-soft}) is even quantitatively comparable to the 
corresponding hard core system. In soft core systems the other possible mode of
stress relaxation, namely, coherent inclusion of one particle each in all
the existing layers is observed at a larger well depth of 
$\mu\approx 16$. We shall discuss about this farther in Sec.\ref{heat} while
discussing heat transport in this soft core system.

\section{Kinetics of layering}

The large hysteresis loops associated with the layering transition
obtained in the last section makes it clear that the kinetics of this
transition is slow. To study the lifetime of the kink-antikink pairs
(surface step), we resort to an MD simulation, using a 
velocity Verlet algorithm\cite{daan}, with the unit of time
given by $\tau = \sqrt{m \sigma^2/k_B T}$, where  
$m (= 1)$ is the mass of the hard disks. 
Using values of $m$ and $\sigma$ typical for atomic systems like Ar or 
Rb, $\tau \approx 1 {\rm ps}$. A time step of 
$\Delta t = 7\times10^{-5} \tau$ 
conserves the total energy to within $1$ in $10^3$ (at worst).

\begin{figure}[t]
\begin{center}
\includegraphics[width=7.0cm]{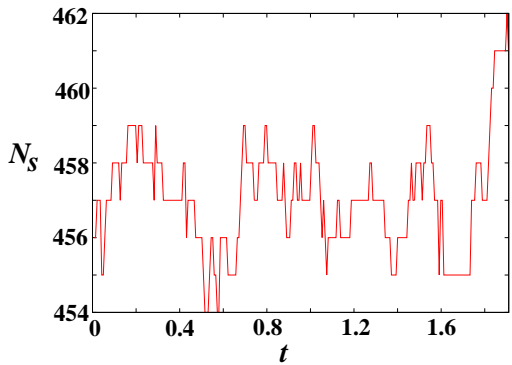}
\end{center}
\caption{Plot of the number of particles in the solid region $N_s$
as a function of time $t$ (in units of $\tau$) clearly shows that 
the displaced particles are pushed back into the region $\cal {S}$.
}
\label{step-annhilate}
\end{figure}

Starting with an equilibrium configuration for the hard disk system,
taken from our MC runs as
discussed in Sec. II at $\mu = 9.6$
corresponding to a $22$-layer solid, we create a unit surface step of
length $l$ by displacing a few interfacial atoms from the solid region
into the liquid and `quench' across the transition to $\mu=4.8$, where
a $21$-layer solid is stable.  The rest of the parameters are kept
identical to those given in section II.  We observe that the
fluctuation thus created rapidly relaxes back and the surface step
vanishes as the atoms are pulled back into the solid. We illustrate this 
by plotting the number of hard disks within the solid region as a function 
of the MD time steps (Fig. ~\ref{step-annhilate}). 
As soon as a step is
created, the line of atoms in the portion of the solid thus exposed
bend to fill up the gap created between the atomic layers and the edge
of the potential trap. This generates considerable local elastic
stress. Also, the liquid layer lying immediately adjacent to the solid
has a lot of orientational and solid-like order. For short times it
responds elastically to the presence of an increased local density of
atoms. The combined elastic response therefore pushes the displaced
atoms back into the solid region thereby annihilating the
step. This annihilation is a transient response and
happens within a time of $\tau$, whereas
the lifetime of the metastable $22$-layered solid is about $10\tau$,
which is one order of magnitude larger!
Indeed, a free energy audit involving a bulk free energy gain
$\Delta F \sim 1/L_s$, going from a 22 to 21 layered solid, and an elastic
energy cost $\sim \log(l)$ for creating a step of size $l$, reveals that a
surface step is stable only if $l\geq l^\ast \sim 1/L_s$. For small
$L_s$, the critical size $l^\ast$ may therefore exceed $L_x$, the total
length of the interface. Of course, if the step spans the entire
length of the interface, there is no bending of the atomic lines and
there is no elastic energy cost. This explains the slow kinetics since
the system has to wait till a rare random fluctuation, which displaces
all the atoms in a solid layer across the interface coherently, gives
rise to layering transition. Although we have
explicitly demonstrated this for the hard disk system, we believe that
similar considerations should be appropriate for the soft disk and
Lennard Jones systems too.

The slow kinetics of the layering transition may have an impact on the
transfer of momentum across the liquid-solid interface in the form of
regular sound waves or acoustic shocks. The large effective
compressibility of the solid at the layering transition as evidenced
by the jump in the density as the chemical potential is increased by
an infinitesimal amount (Fig.~\ref{hyst}) should reduce the velocity of
sound considerably. The propagation and scattering of sound in an
inhomogeneous region with coexisting phases has been studied
extensively \cite{lanlif2,zener,isakovich,onuki} in the past. The
transfer of mass between coexisting phases at inter-phase boundaries
is known to slow down and dissipate sound waves traveling through the
system. Our system has an artificially created inhomogeneity, which
should have a similar effect on its acoustic properties.

Further, the mechanism of stress relaxation of a thin ($L_s$ small)
solid via the transfer of an entire layer of atoms may be exploited
in a variety of practical applications, provided we can eject this
layer of atoms deep into the adjoining fluid and enhance its lifetime.
We may be able to use the ejected layer of atoms to create monolayer
atomic films or coatings\cite{sakhi}. Highly stressed mono-atomic 
layers tend to
disintegrate or curl up \cite{novoselov} as they separate off from the
parent crystal. It may be possible to bypass this eventuality, if the
time scale of separation is made much smaller than the lifetime of the
layer. Can acoustic spallation \cite{zeldovich} be used to cleave
atomic layers from a metastable, stressed nanocrystal?

In the next section, therefore, we study the response of the liquid
solid interface in our system of hard disks to acoustic shocks with a
view to studying the effect of the layering transition on acoustic
shock propagation and dissipation as well as the properties of the 
ejected layer.

\section{Mass and Momentum transfer}
\begin{figure}[t]
\begin{center}
\includegraphics[width=8.0cm]{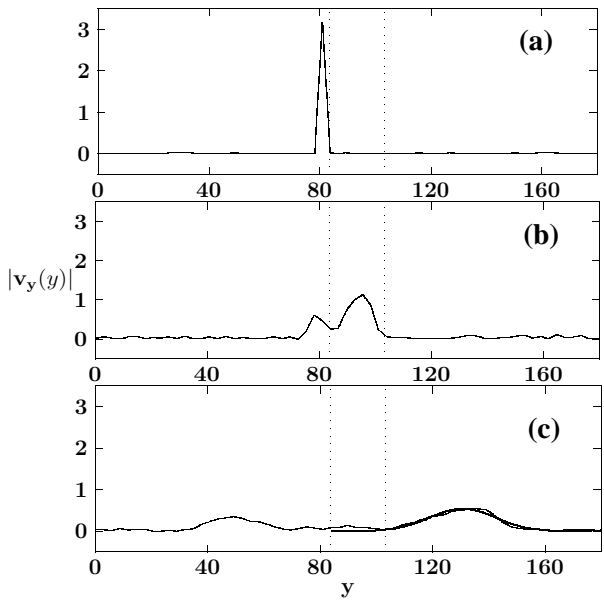}%{fig3a-bw.eps}
\end{center}
\caption{
(a)-(c) Plot of the absolute value of the momentum $|v_y(y)|$ for molecular
dynamics times $t = 0.0007$ (a), $0.2828$ (b) and $2.8284$ (c). The 
dotted lines show the position of the solid-fluid interfaces. 
The fit to a Gaussian (thick solid line) is also shown in (c).  
Curves such as in (a)-(c) are obtained by averaging 
over $100-300$ separate runs using different realizations of the initial
momentum.}
\label{pulse}
\end{figure}

Consider sending 
in a sharp laser (or ultrasonic) pulse, producing
a momentum impulse ($v_y(t=0) = V_0$) over a thin region in $y$ spanning the
length $L_x$ of the simulation cell. This results in a weak acoustic 
shock \cite{zeldovich} (corresponding to a laser power $\approx 10^2$ mW and a 
pulse duration $1 {\rm ps}$ for a typical atomic system). 
The initial momentum pulse travels through the solid and 
emerges at the far end (Fig.~\ref{pulse}) as a broadened Gaussian. The  
width of this Gaussian pulse, $\Delta$, is a measure of absorption of the 
acoustic energy of the pulse due to combined dissipation in the liquid, 
the solid and at the interfaces \cite{lanlif2,zener,cahill}. 
For large enough pulse strengths $V_0$, this is accompanied by
\noindent
a {\it coherent} ejection of the (single) outer layer of atoms into the fluid.  
Note that in our MD simulations, to reduce interference from the reflected 
pulse through periodic boundary conditions, we increase the fluid regions on 
either side, so that for the MD calculations we have a cell of size 
$22.78\times186.98$ comprising $3600$ particles. This is accomplished by 
separately equilibrating two liquid regions of appropriate size and density
and smoothly sandwiching our equilibrated system (which includes 
the solid strip) in between these liquid regions. The whole $3600$ particle
system is equilibrated for a further 
$10^4$ MCS before it is used as an initial 
condition for the MD simulations. In Fig.~\ref{pulse} we show the initial 
momentum pulse with strength $V_0 = 6.$ as produced within a narrow strip 
of size $\sim \sigma$, just to the left of the solid region and the curves are
fitted to a Gaussian (and the width $\Delta^2$ extracted) when the maximum of 
the pulse reaches a fixed distance of $44.1$ from the source. A reflected pulse can also be seen. 

\begin{figure}[t]
\begin{center}
\includegraphics[width=5.5 cm]{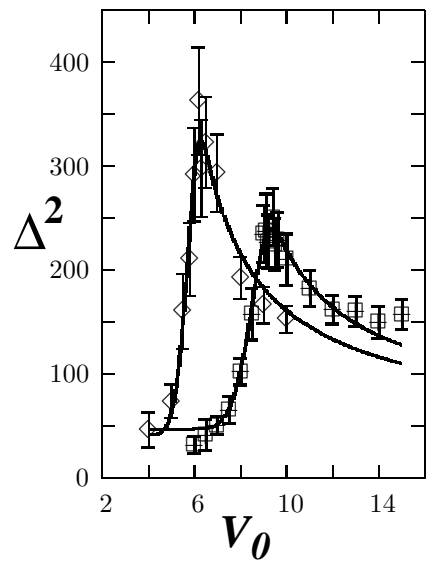}
\end{center}
\caption{Plot of the squared width $\Delta^2$ of the momentum 
pulse after it emerges from the solid as a function 
of $V_0$ for $\mu = 4.8\,(\,\Diamond)$ and $9.6\,(\,\Box)$. The 
solid lines are fits to an effective liquid theory. 
The peak in $\Delta^2(V_0)$ so produced is more prominent for 
the metastable $22$ layered solid $\mu = 4.8 $ than for the 
stable ($\mu = 9.6$) system showing a more coherent momentum 
transfer in the former case.}
\label{reson}
\end{figure}
\begin{figure}[t]
\includegraphics[width=8.0cm]{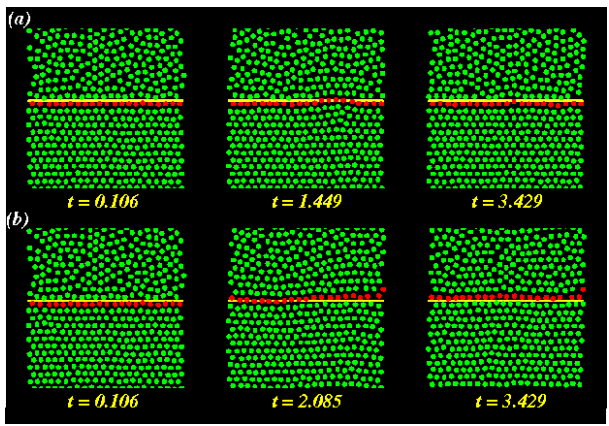}
\caption{(Color online)
(a) Configuration snapshot from a portion of our MD cell showing 
hard disk atoms (green circles) at the solid (bottom)- liquid 
interface (yellow line) as a weak momentum pulse ($V_0 = 2$) emerges 
into the liquid, at three different times, for $\mu = 4.8$. The pulse 
initially ejects a few atoms from the solid but are subsequently pulled 
back due to the large elastic strain cost in bending
of the interfacial crystalline layer (red circles). 
(b) The same for a stronger momentum pulse, $V_0 = 6$. This time the pulse
strength is sufficient to eject the layer.
}
\label{movie}
\end{figure}

When a shock wave, which propagates through a conventional solid
emerges from the free surface, the compressed material expands 
-- or unloads -- to zero pressure \cite{zeldovich}. The unloading 
(rarefaction) wave travels backwards into the
material with the speed of sound. The response of the solid
depends on the specific nature of the shock front. For a shock wave
with an approximately Gaussian profile as in our case, significant
negative pressures can develop at the interface where the shock
emerges due to the interaction of the forward and the reflected waves
and a portion of the solid may split off by a process known as ``spallation". 
Spallation in bulk solids like steel needs acoustic
pressures in excess of $10^5$ N/cm$^2$ \cite{zeldovich} usually
available only during impulsive loading conditions; the ejected layer
is a ``chunk'' of the surface. In contrast, the pressures generated by
the shock wave in our system causing coherence {\it nanospallation} involves 
much smaller surface stresses of the order $k_B T/\sigma^2 \approx  
10^{-5}$ N/cm$^2$.  This difference comes
about because unlike a bulk system, a strained nanocrystal on the
verge of a transition from a metastable $n+1$ to a $n$ layered state
readily absorbs kinetic energy from the pulse. Also, as mentioned before, 
a confined solid strip in two-dimensions has very strong smectic ordering
which effectively decreases the coupling between the 
layers\cite{andrea, debc}. In fact, a quasi one-dimensional solid 
($L_x \gg L_s$) is better regarded as a smectic with weak solid like 
modulations. The fact that surface indentations are unstable 
(Fig.~\ref{movie}) unless of a size comparable to the length of 
the crystal, $L_x$, ensures that a full atomic layer is evicted almost 
always, leading to coherent absorption of the pulse energy.  
The coherence of this absorption 
mechanism is markedly evident in a plot of $\Delta^2$ against $V_0$
which shows a sharp peak (Fig.~\ref{reson}). Among the two systems studied by 
us, {\it viz.}, a metastable ($\mu = 4.8$) and a stable ($\mu = 9.6$)  $22$ 
layered solid, the former shows a sharper resonance. 
Note that the absorption of momentum 
is largest when the available kinetic energy of the pulse exactly 
matches the potential energy required to eject a layer. To elucidate this 
fact further, we plot the configurations of the metastable system 
($\mu = 4.8$) as the pulse travels through the system, for two different
pulse strengths, $V_0 = 2$ and $V_0 = 6$ (Fig.~\ref{movie}). The weaker
momentum pulse ($V_0 = 2$) initially ejects a few atoms of the interfacial
crystalline layer of the metastable $22$ layered
%\clearpage 
%\clearpage 
solid. However, the resulting large non-uniform elastic strain 
evidenced by the bending of lattice layers causes these atoms 
to be subsequently pulled back into the solid. This effect is the same
as that seen in last section. Only a stronger pulse,
$V_0 = 6$, capable of ejecting a complete lattice layer succeeds in reducing 
the number of solid layers by one leading to an overall lower elastic energy. 

The eviction of the atomic layer is 
therefore assisted by the strain induced interlayer transition and 
metastability of the $22$ layered solid discussed above. Spallation
is also facilitated if the atomic interactions are anisotropic so that
attraction within layers is stronger than between layers (eg. graphite
and layered oxides \cite{novoselov}), for our model of purely repulsive
hard disk solid, an effective intralayer attractive potential of mean force is 
induced by the external potential\cite{debc}.
 
\begin{figure}[t]
\begin{center}
\includegraphics[width=5.0cm]{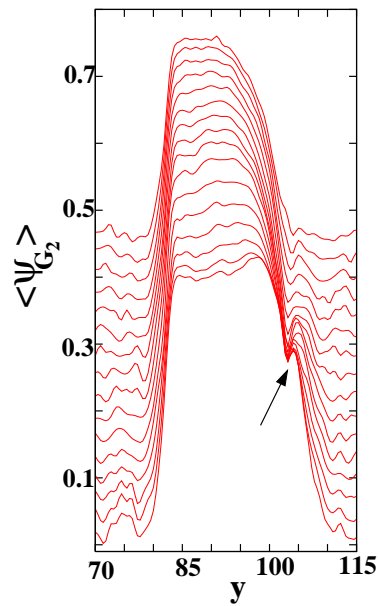}%{Figures/spalitza-fig16.eps}
\end{center}
\caption{(Color online)
A plot of the time development $\la \psi_{{\bf G}_2}(y)\ra$.
The solid ejects a layer (shown by an arrow) which subsequently dissolves in 
the fluid. The curves from bottom to top correspond to time slices 
at intervals of $\Delta t = .07\tau$ starting from $t = 1.06\tau$ (bottom). 
We have shifted each curve upward by $.03 t/\Delta t$ for 
better visibility. }
\label{life}
\end{figure}

The spallated solid layer emerges from the solid surface into the fluid, and 
travels a distance close to the mean free path; whereupon it disintegrates due 
to viscous dissipation (Fig.~\ref{life}). 
%\clearpage
%\clearpage
A simple estimation of the life-time of the ejected layer may be undertaken
as follows.
To obtain the life time of the spallated layer we obtain the time 
development of the Fourier component 
of the local density correlation $\la \psi_{{\bf G}_2}(y)\ra$ 
which is just a time dependent generalization of the quantity 
shown in Fig.\ref{g1g2}. We obtain this by 
averaging, at each time slice $t$, the quantity 
$\sum_{\la i,j\ra} \exp(-i\,{\bf G}_2\,\cdot\,({\bf r_i - r_j}))$ 
over all nearest neighbour pairs $\la i,j\ra$ with centre of the
vector $({\bf r_i - r_j})$ lieing 
within a strip of width $\sim \sigma$ centered 
about $y$ and spanning the system in $x$. The wavenumber 
${\bf G}_2 = (2 \pi/d) {\bf \hat{n}}$ where
$d = .92$ is the distance between crystal lines in the 
direction ${\bf \hat{n}}$ normal to the fluid-solid interface. The 
solid (central region with $\psi_{{\bf G}_k}(y,t) \ne 0$, for $k=1,2,3$) 
ejects a layer (shown by 
an arrow in Fig.~\ref{life}) which subsequently dissolves in the fluid.
The lifetime of the layer is around 
2-3 time units ($\tau$) which corresponds to a few ps for typical atomic
systems. The lifetime increases with decreasing viscosity of the surrounding 
fluid. Using the Enskog approximation \cite{chapman} to the hard disk 
viscosity, one can calculate the bulk viscosity for a hard disk fluid to be
\cite{chapman,wood}
\begin{eqnarray}
\zeta_E = \frac{16}{\pi}\zeta_{00}\eta^2g(\sigma)
\end{eqnarray}
where $\zeta_{00}$ is a constant and $g(\sigma)$ is the pair-correlation
function at contact.
For a system of hard disks with $m = \sigma = \beta = 1$
$\zeta_{00} = 1/2\sqrt\pi$\cite{wood}. Thus, $\zeta_E \propto \eta^2$ and 
we estimate that by lowering the fluid density one may increase 
the lifetime of the layer considerably. The lifetime enhancement is 
even greater if the fluid in contact is a low density gas (when the 
interparticle potential has an attractive part \cite{LJ-visc}).

\subsection{Effective liquid theory}

\begin{figure}[t]
\begin{center}
\includegraphics[width=7.0cm]{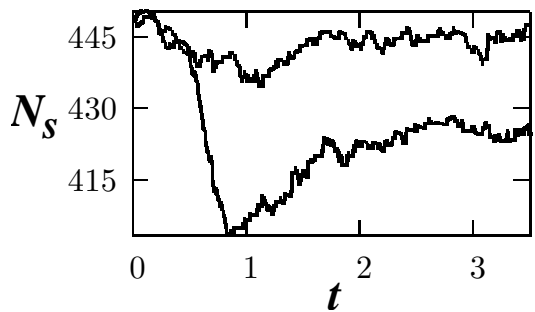}
\end{center}
\caption{A plot of the total number of particles $N_s$ within the solid region 
($\mu = 4.8$) as a function of time for $V_0 = 1$ (top) and $6$.
Note oscillations in $N_s$; only the stronger 
pulse changes the number of solid layers from $22$ to $21$. } 
\label{nst}
\end{figure}

\begin{figure}[t]
\begin{center}
\includegraphics[width=7.0cm]{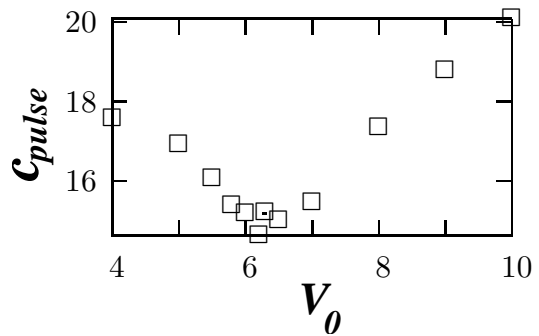}
\end{center}
\caption{Average pulse velocity $c_{pulse}(V_0)$ for $\mu = 4.8$; note the dip 
in $c_{pulse}$ where absorption is strongest.} 
\label{pdec}
\end{figure}

The absorption line~-shape may be understood within a phenomenological 
``effective liquid'' approximation. The extra 
absorption producing the prominent peak in Fig.~\ref{reson} is due to the 
loss of a whole layer from the solid into the liquid (Fig.~\ref{movie}). For 
small $V_0$, the confined solid responds by centre of mass fluctuations 
($q\to 0$ phonons) shown by oscillations of $N_s$ with time (Fig.~\ref{nst}).
\begin{figure}[h]
\begin{center}
\includegraphics[width=6.0cm]{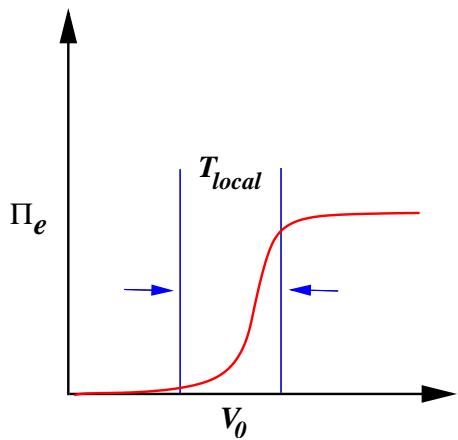}
\end{center}
\caption{(Color online)
A schematic diagram showing the momentum transfer as assumed in our
phenomenological theory.} 
\label{mom-theory}
\end{figure}
Scattering from this and other sources \cite{cahill,lanlif2,zener,isakovich,
onuki} constitute a background which we ignore, as a first approximation, for 
simplicity. Within our approximation, the momentum loss 
at the interface is modelled as 
regular dissipation within a liquid strip of (fictitious) width $\xi$. The
expected momentum transfer at the interface $\Pi_e = \Pi_0 \times$ the 
probability that the momentum $\Pi_0$ required to eject the layer, exists
(see Fig.~\ref{mom-theory}).
If a local ``temperature'' $T_{local}$ measures the degree of (de)coherence 
of the momentum transfer, then $ \Pi_e = (1/2) \Pi_0 
{\rm erfc}[(\Pi_0 - V_0)/\sqrt{2 k_B T_{local}}]$ and $\xi$ 
may be extracted from $V_0 - \Pi_e = V_0 \exp(-\alpha \omega^2 \xi)$.
Substituting for $\Pi_e$ we obtain the extra absorption due to the
interface,
\begin{equation}
\Delta^2 =  4.\alpha c_0^2 \xi = - a \log [ 1 - \frac{\Pi_0}{2 V_0}{\rm erfc}\{\frac{\Pi_0 - V_0}{\sqrt{2 k_B T_{local}}}\} ]
\label{ela}
\end{equation}
We use $a,\Pi_0$ and $T_{local}$ as fitting parameters. In Fig.~\ref{reson} we 
show a fit to Eq. \ref{ela} of our MD data and observe that it reproduces all
the features remarkably well. The larger error-bars near the peak 
in $\Delta^2(V_0)$ reflects the difficulty of fitting a Gaussian to the 
transmitted pulse when dissipation is large. Indeed, in this region the
pulse shape is systematically distorted away from Gaussian due to effects 
beyond the scope of our simple theory. Large fluctuations in $N_s$ 
(Fig.~\ref{nst}) lead to expected \cite{lanlif2,zener,isakovich,onuki} and  
detectable decrease in average pulse speed (Fig.~\ref{pdec}).

\section{Heat transport}
\label{heat}
In this section we focus on the transport of energy across the system of
forced solid in contact with its own liquid. The heat conductivity
$\l$ is defined by the celebrated Fourier's law
${\bf j}_E = -\lambda{\bf  \nabla {T}}$ 
where ${\bf j}_E$ is the heat current density and ${\bf \nabla {T}}$ is the
temperature gradient. The transport of heat through small and low dimensional
systems has enormous significance in the context of designing useful
nano-structures \cite{cahill}. A large number of recent studies in lower
dimensions has shown that heat conductivity is infact divergent 
as a function of system size\cite{livi,bonet,grass}. Thus it is more
sensible to calculate the heat current or conductance of the system directly,
rather than the heat conductivity.
Therefore we focus on the heat current densities $j_E\,(\,>0\,)$ 
flowing across the solid-liquid interface from high to low temperature
and heat conductance $G = j_E/\D T$
(or resistance $R=1/G$),  $\D T \,(\,>0\,)$ being the temperature difference 
between the two edges of a given region.  In this study, we
are particularly interested in exploring the impact of structural changes,
viz. the layering transitions, on heat transport. This is to remember that,
layering transitions are strongly dependent on the small system sizes and 
gets washed away as one goes to larger systems. Recently, electrical
\cite{myecond} and thermal transport\cite{debc-heat} studies on confined
solid strips have revealed strong signatures of structural transitions
due to imposed external strain.
Heat transport across a model liquid-solid
interface has been studied in three dimensions with the interatomic potential
being Lennard-Jones\cite{barrat}. In Ref.\cite{barrat} it is shown that the 
Kapitza resistance\cite{kapitza}, the interfacial resistance, can reach 
appreciably large values when the liquid does not wet the solid.

\subsection{Non-equilibrium molecular dynamics} 
The specific context in which we study thermal properties of the
liquid-solid interface is the same as that we have used for our
studies of the acoustic properties in the last section. Again, a solid
region is created within a liquid using an external chemical potential
trap. 
We report results for $1200$ particles interacting via the soft disk potential
$u(r_{ij})=1/r_{ij}^{12}$ taken within an area of $24\times 60$.
In absence of any external potential, a 2d system of soft disks
at this density $\r\approx0.65$ remains in the fluid phase.
The length scale is set by  soft disk diameter $d=1$,
energy scale by temperature $\kb T$ and the time scale by
$\tau_s=\sqrt{m d^2/\kb T}$. The unit of energy flux $j_E$
is thus set by $(\kb T/\tau_s d)$. The unit of resistance and
conductance are $\tau_s d$ and $(\tau_s d)^{-1}$ respectively.
The smooth interaction potential allows us
to use standard MD simulations with a velocity Verlet algorithm. The
time step of $\delta t = 10^{-3}\,\tau_s$ in our MD ensures that the total
energy is conserved (in equilibrium) to within $10^{-4}$. Periodic 
boundary conditions are applied in the $x$-direction.  We use the standard
velocity Verlet scheme of MD with equal time update of time-step $\d t$,
except when the particles collide with the `hard walled' heat reservoirs at 
$y=0$ and $y=L_y$. We treat the collision between the particles and the
reservoir as that between a hard disk of unit diameter colliding
against a hard structure less wall. If the time, $\tau_c$, of the next
collision with any of the two reservoirs at either end is smaller than
$\delta t$, the usual update time step of the MD simulation, we update
the system with $\tau_c$. During the collision with the walls Maxwell
boundary conditions are imposed to simulate the velocity of an atom
emerging out of a reservoir at temperatures ${T}_L$ (at $y=0$) or ${T}_R$
(at $y=L_y$)\cite{bonet}.  
This means
that whenever a soft disk collides with  either the left or the right wall 
it gets reflected back into the system with a velocity chosen from 
the distribution
\begin{equation}
f(\vec v)=\frac{1}{\sqrt{2\pi}}\left(\frac{m}{k_B T_W}\right)^{3/2}
|v_y|\exp\left(-\frac{m {\vec v}^2}{2k_B T_W}\right) 
\end{equation}
where $T_W$ is the temperature ($T_L$ or $T_R$) of the wall on which
the collision occurs.  
During each collision energy is exchanged between the system and the
bath. Thus in our molecular dynamics simulation, the average heat
current flowing through the system can be found easily by computing the
net heat loss from the system to the two baths  (say $Q_L$
and $Q_R$ respectively) during a  large time interval $\tau$, once
the system has reached steady state. 
The steady state heat current from right to left bath is given by 
$ \la J \ra
= \lim_{\tau \to \infty} Q_L/\tau = -\lim_{\tau \to \infty} Q_R /\tau$.    
In the steady state the heat current (the heat flux density integrated over
$x$) is independent of $y$. This is a requirement coming from current
conservation. For a homogeneous system $j_E=\la J \ra/L_x$.
However if the system has inhomogeneities then the flux
density itself can have a spatial dependence and in general we can
have $j_E=j_E(x,y)$. In our simulations we have looked at $j_E(x,0)$ 
and $j_E(x,L_y)$.

\subsection{Results}
\begin{figure}[t]
\begin{center}
\includegraphics[width=8.6cm]{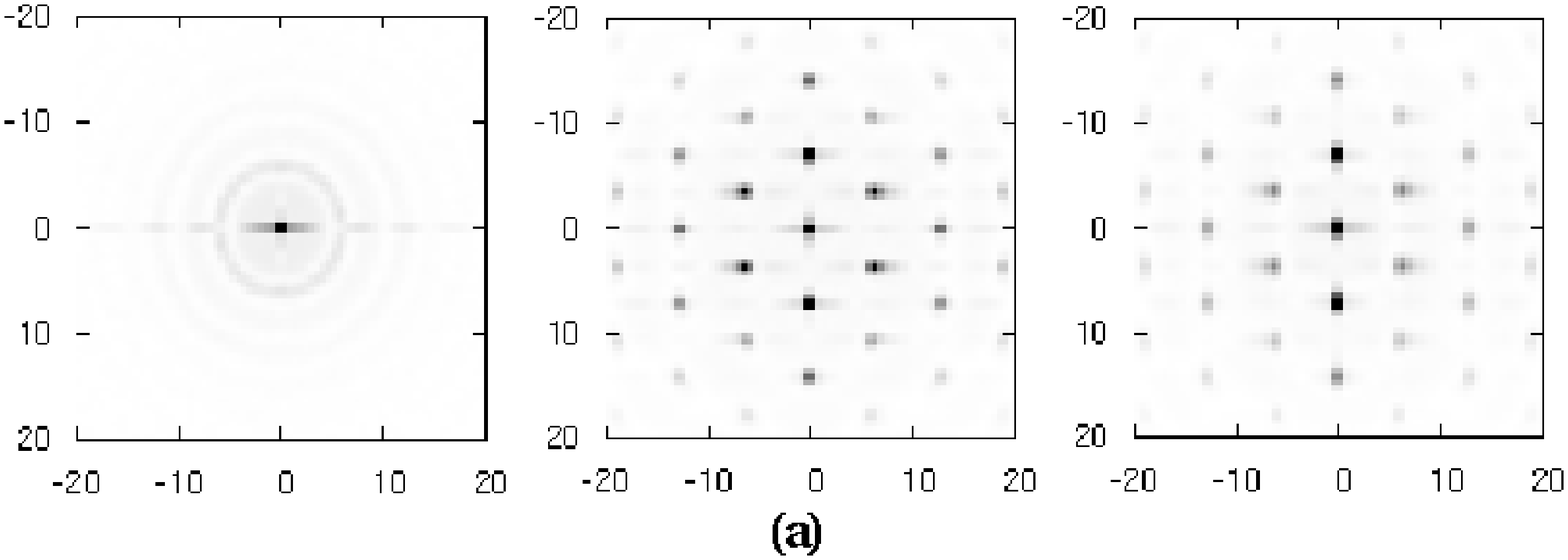}%{heat-sks.eps}
\vskip 1cm
\includegraphics[width=4.cm]{dens-prof.eps}
\hskip .2cm
\includegraphics[width=4.cm]{comprs-prof.eps}
\end{center}
\caption{  
(a) From left to right, structure factors in liquid, solid and interface 
regions. Interface shows clear smectic profile. Data taken at $\mu=13$.
(b) The local density profile along $y$-direction at $\mu=13$.
(c) The isothermal compressibility $\kappa_T$ as a function of $y$ at $\mu=13$.
Compressibility shows strong peaks near the interfaces. Due to small size, 
interfacial enhancement of compressibility permeates right through the whole 
of solid region. In (b) and (c) lines are guides to eye.
}
\label{heat-struct}
\end{figure}
\begin{figure}[t]
\begin{center}
\includegraphics[width=4.cm]{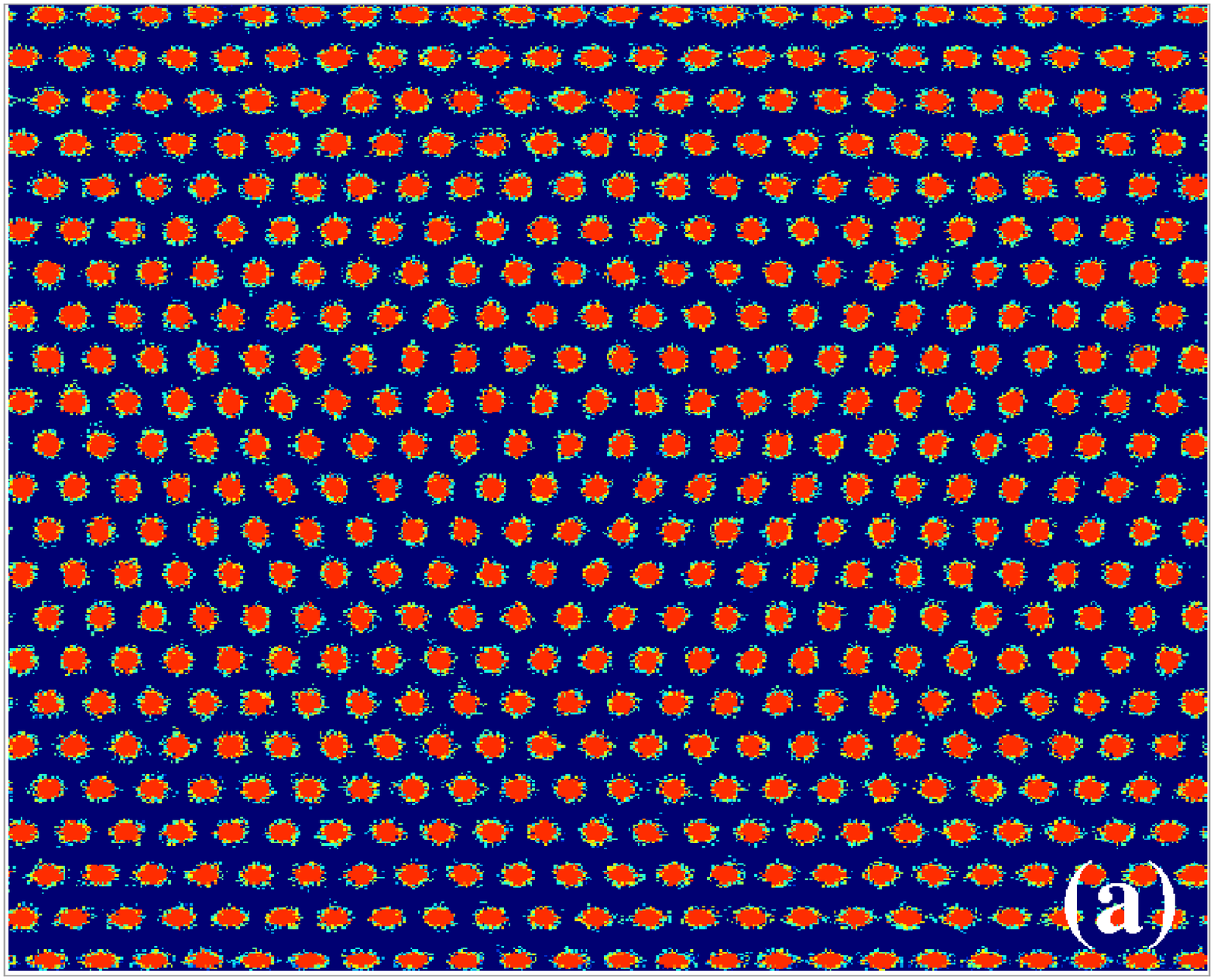}%{disl-sol-24.eps}
\hskip .2cm
\includegraphics[width=4.cm]{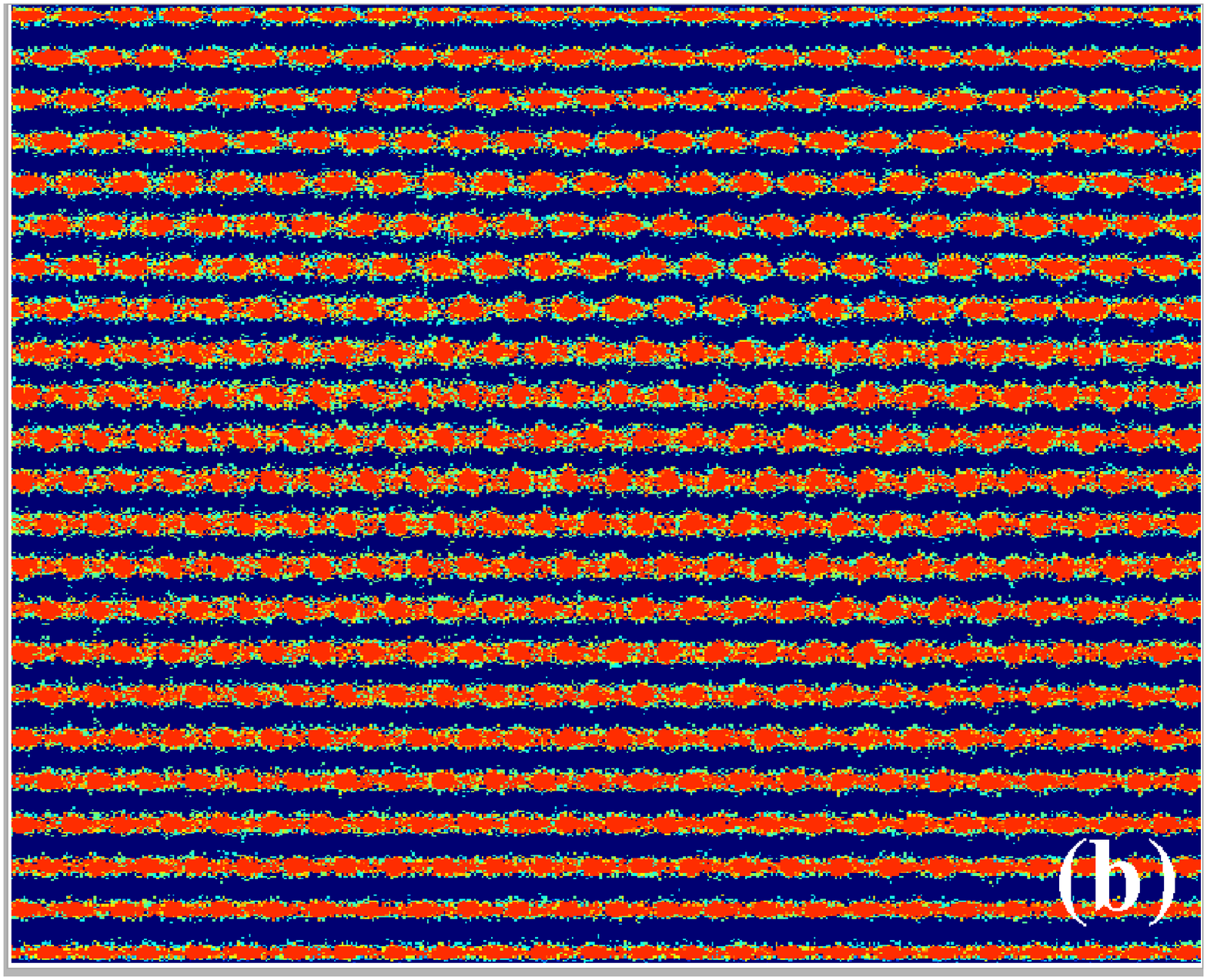}%{sld-24.eps}
\vskip .1cm
\includegraphics[width=4.cm]{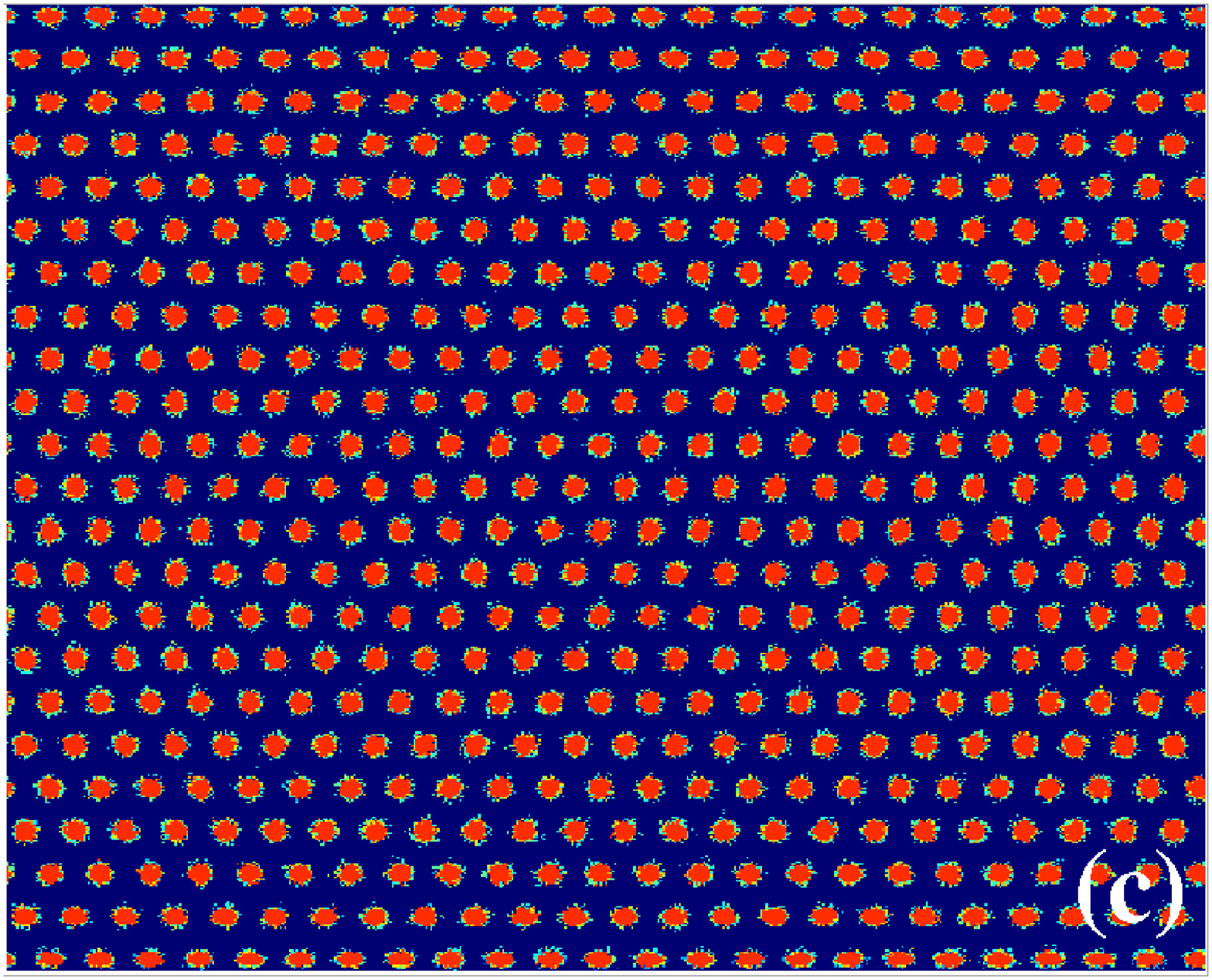}%{dislo.eps}%{disl-sol-24.eps}
\hskip .2cm
\includegraphics[width=4.cm]{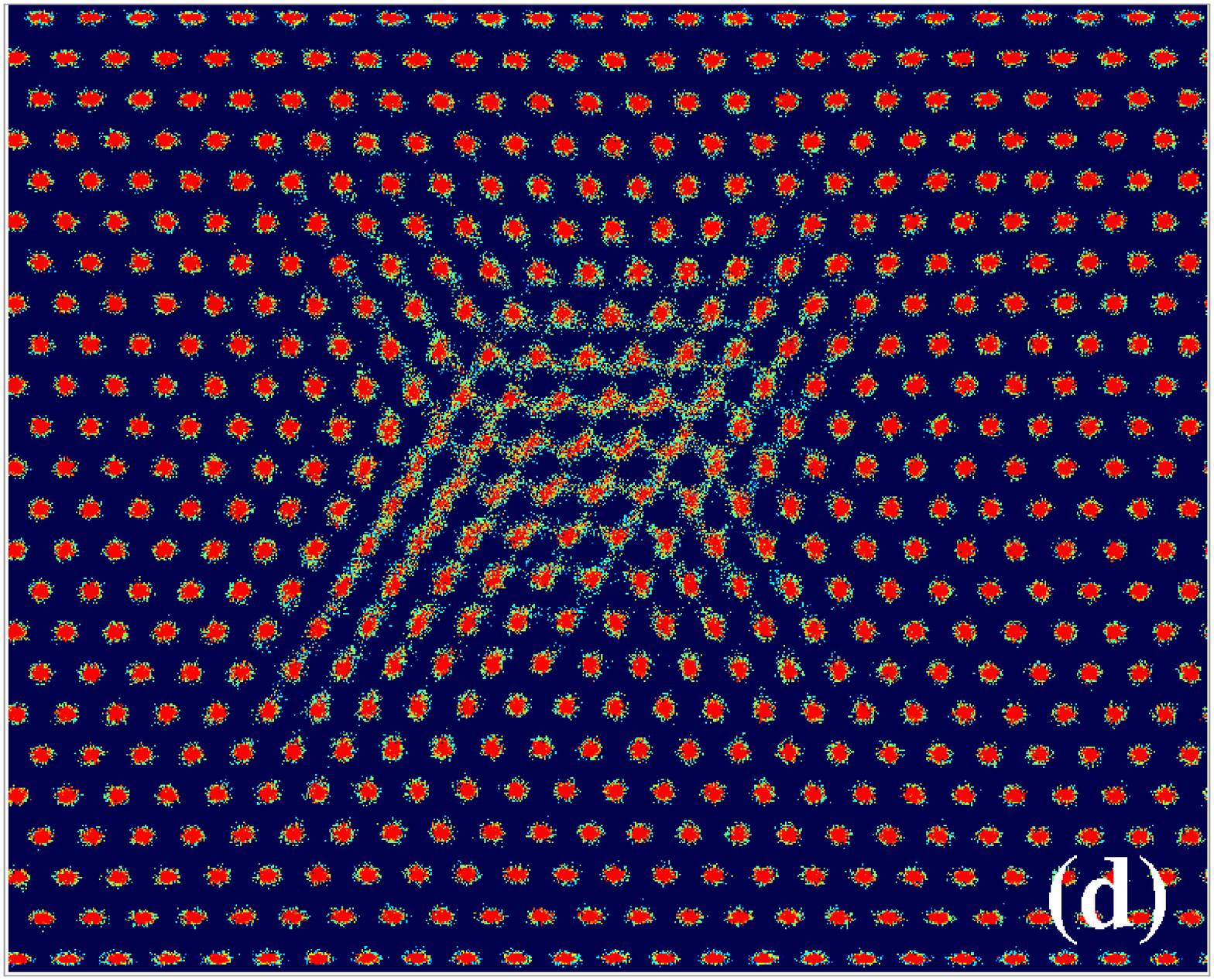}%{solid-mu24.eps}%{sld-24.eps}
\end{center}
\caption{(Color online)
Overlapped density plot of $500$ configurations in the region trapped
by external potential $\mu$: 
(a) A $23\times 23$ triangular lattice solid at $\mu=8$.
(b) Local density peaks hop in $x$-direction to incorporate $>23$ particles
in lattice planes in response to increased potential $\mu=11$.
(c) A $24\times 23$ triangular lattice solid at $\mu=12$. Notice the 
increase in particle numbers in the lattice planes.
(d) Configuration  obtained after $15000 \d t$ as a $24\times 23$  
steady state solid at $\mu=16$ is quenched to $\mu=24$. This shows a
dislocation pair -- a $23$-layered region trapped in between a $24$-layered
solid. At steady state (after a time $10^5 \d t$) dislocations annihilate
to produce a $24\times 24$ triangular lattice solid.
Color code: blue (dark): low density and red (light): high density.
}
\label{dislo}
\end{figure}

In Fig.\ref{heat-struct} we show that even in presence of a temperature
difference across the system, the solid-liquid interface is formed and the
liquid near the interface shows smectic-like density modulation due to
the presence of nearby solid. These features are apparent from the structure
factor calculated outside the region ${\cal S}$ (liquid), inside the region
${\cal S}$ and in the small region over which external potential goes to the 
value $-\mu$ from zero (Fig.\ref{heat-struct}.(a)). The local density profile
also shows constant large value corresponding to the solid formed in region 
${\cal S}$  (Fig.\ref{heat-struct}.(b)). The density of the liquid near the
cold ($\kb T=0.5$) left-reservoir is higher than the density of  liquid near
the hot ($\kb T=1.5$) right-reservoir. In  Fig.\ref{heat-struct}(c) we plot 
the local compressibility $\kappa_T(y)$ defined via 
$\kappa_T=\r^{-2}(\p\r/\p\mu)_T$. 
The compressibility of the interfaces is very large making 
the narrow solid region also unusually compressible pointing to the presence 
of large local number fluctuation. This behaviour helped in large shock
absorption as discussed in Sec.V.
Before going into the details of heat transport in this system,
let us first enlist some details of the structural transitions obtained. 
With increase in the strength of the trapping potential,
both in equilibrium and in non-equilibrium situations, we observe two modes of
density enhancement: (a) A whole layer of particles enter to increase the
number of lattice planes in $y$-direction. This happens, e.g., as $\mu$ is 
increased from $7$ to $8$. Thus in this mode the
inter lattice plane separation decreases (see Fig.\ref{dislo}(a)). 
(b) Each of the lattice planes 
grow by one atom thereby decreasing the interatomic separation within
a lattice plane. This happens, e.g., as one increases $\mu$ from $10$ to $12$ 
(see Fig.\ref{dislo}(c)). In the intermediate configurations one observes
metastable dislocation pairs ( Fig.\ref{dislo}(d)) and peaks in the local
particle density that hops back and forth between two neighbouring 
positions ( Fig.\ref{dislo}(b)) to maintain commensurability.
In the mode (a) one observes a sudden compression 
in the $y$-direction associated with positive $\ve_d$ and $\g_d$, while in 
the mode (b) the system undergoes tension associated with negative $\ve_d$ 
and $\g_d$. With increase in $\mu$, these two modes
repeat one after another, in cycle. This allows the system to release the
extra stress developed in one direction due to particle inclusion in 
the previous cycle  by inclusion of particle in the other direction in the 
next one. Certainly, at large enough $\mu$
the solid region goes towards very high packing fraction, thereafter
stopping the process of particle accumulation. To summarize the major
structural changes obtained in soft disk solid, we find, strained
triangular solids with $23\times23$, $24\times23$, $24\times24$ unit cells 
at $\mu=8,\,12$ and $24$ respectively (See Fig.\ref{dislo}).
The large elastic and core energy cost inhibits the formation of an equilibrium
nano solid with dislocations -- even if dislocations form, the solid
eventually gets rid of them by either incorporating particles from or 
ejecting them to the liquid part. 

\begin{figure}[t]
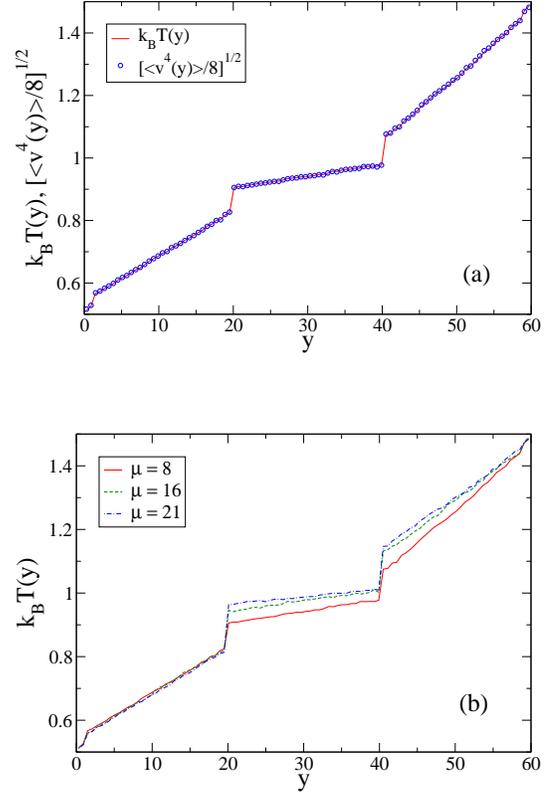

\begin{center}
\includegraphics[width=7cm]{LTE.eps}%{lte-temp.eps}%{Figures/spalitza-fig20.eps}
\vskip 1.cm
\includegraphics[width=7cm]{tprof.eps}%{Figures/spalitza-fig20.eps}
\end{center}
\caption{(Color online)
(a) Plot of temperature profile $\langle 1/2~mv^2(y)\rangle$ 
and $m\sqrt{\langle v^4(y) \rangle/8}$ at $\mu=8$.
(b) The temperature profile $\langle 1/2~mv^2(y)\rangle$ as a function of $y$, 
the system coordinate perpendicular to the reservoirs, for 
$\mu=8,\,16,\,21$.
}
\label{lte}
\end{figure}

Before any measurement is done,
the system is allowed to reach the steady state where the current 
density integrated over the whole $x$-range is
the same at all $y$. 
If LTE is maintained, $v(y)$ is expected to obey Gaussian distribution
locally. That gives definitions of local temperature from all the even moments
of $v(y)$. Thus  $\kb T(y) = \langle 1/2~mv^2(y)\rangle$ and 
$\kb T(y) = m\sqrt{\langle v^4(y)\rangle/8}$ etc.
To check for the local thermal equilibrium (LTE) from our simulation,
we find $\la v^2(y)\ra$ and $\langle v^4(y) \rangle$ as a function of
distance $y$ from cold to hot reservoir and compare the above mentioned 
definitions of $\kb T(y)$(Fig.~\ref{lte}(a)). 
From Fig.~\ref{lte}(a)
it is evident that the temperature profile is almost linear in the
single phase regions, like the liquid and the solid, with sharp increase
near the interfaces and the LTE is approximately valid in all regions.
In Fig.~\ref{lte}(b) we plot temperature profiles $\kb T(y)$ as obtained
from $\la 1/2\,m v^2(y)\ra$ at well separated
trapping potentials $\mu=8,~16,~21$. With increased trapping strength,
the temperature difference between the edges of the solid region decreases
indicating an enhancement of heat conductance within the solid.
The temparature jumps at the interfaces also increase with increasing
trapping potential.
Such a jump in the temperature is known as the Kapitza or contact
resistance ($R_K$) \cite{kapitza}. This is defined as,
\begin{eqnarray}
R_K = \frac{\Delta {T}}{j_E}
\end{eqnarray}
where $\Delta {T}$ is the difference in temperature across the
interface. It is evident that the interfaces are the regions of the
highest resistance in the system. This large resitance can be traced
back to large density mismatch at the contact of two phases. 
The conductance of the high temperature
liquid near the right reservoir with $\kb T_R=1.5$ is lower than the low
temperature liquid in the other side in contact with the reservoir with
$\kb T_L=0.5$. The temperatures are expressed in units of $\kb T$. The
heat current, as expected, flows from the right reservoir to the left reservoir.
\begin{figure}[t]
\begin{center}
\includegraphics[width=7.cm]{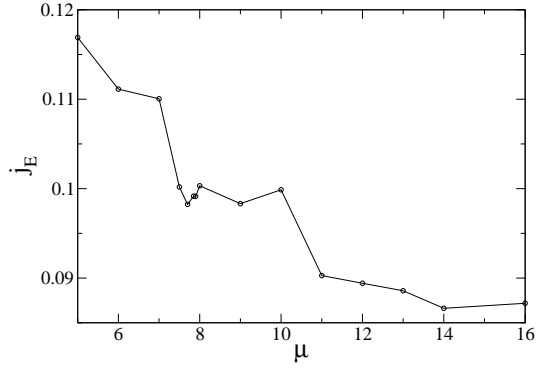}%{jE-new.eps}%{Figures/spalitza-fig22.eps}
\end{center}
\caption{Plot of the heat flux, $j_E$, as a function of the trap depth, $\mu$.
Note that the overall flux decreases as a function of $\mu$. $j_E$ is  expressed
in units of $\kb T/\tau_s d$.
}
\label{flux}
\end{figure}

\begin{figure}[t]
\begin{center}
\includegraphics[width=7.cm]{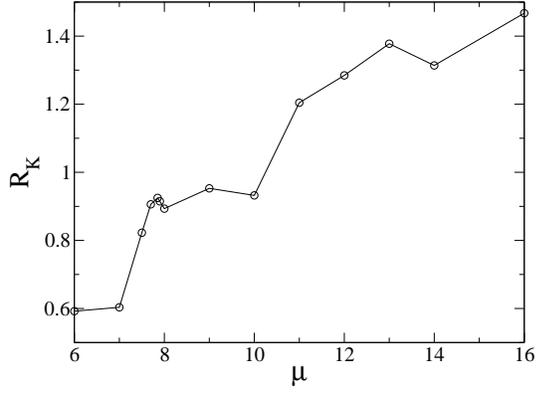}%{kapr-new.eps} %{Figures/spalitza-fig21.eps}
\end{center}
\caption{Plot of the Kapitza resistance, $R_K$, expressed in units of 
$\tau_s d$ as a function of $\mu$, shows a jump at the layering transition. 
}
\label{kapr}
\end{figure}

In Fig.~\ref{flux} we
have plotted the heat flux through the system as a function of
$\mu$. As $\mu$ increases, the atoms from the surrounding liquid get
attracted into the potential well and the density of the solid 
progressively becomes higher at the cost of the liquid. Because of
Kirchoff's law, the heat conductance of a composite system of liquid-solid-liquid connected in series through interfacial regions is dictated by the 
low conductance regions. The decrease in liquid density decreses the pressure
in the liquid regions, thereby reducing the heat conductance in 
them\cite{luding}. The conductance in liquid is always lower than solid.
These result in an overall decrease in the heat flux and consequently the 
overall conductance.
Moreover, at larger $\mu$ there occurs larger density mismatch at the 
interfaces leading to larger Kapitza resistance (Fig.\ref{kapr}).  
The change in density is sharper near the two
layering transitions -- thus  heat flux shows sharper drops  near the
transitions at $\mu=7-8$ and $\mu=11-12$.   
However, close to the layering transition at $\mu \sim
8$ there is a local peak in the value of the heat flux suggesting that
a significant amount of kinetic energy is exchanged between the liquid
and solid through the interface at the layering transition. 
This excess conduction is due to an enhanced number fluctuation in the
direction of heat flow in this mode of layering transition. 

\begin{figure}[t]
\begin{center}
\includegraphics[width=7cm]{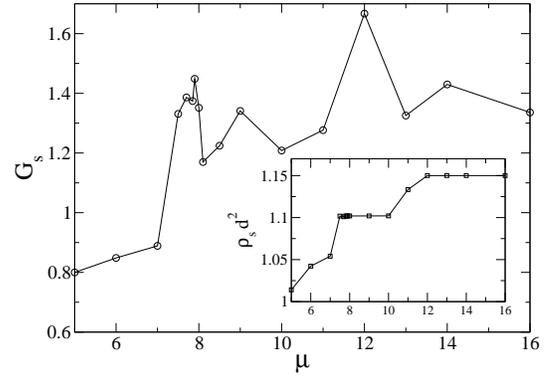}
\end{center}
\caption{Plot of the thermal conductance of the solid region, $G_s$ as a 
function of $\mu$. $G_s$ is plotted in units of $(\tau_s d)^{-1}$. The
inset shows change in solid density $\r_s d^2$ as a function of $\mu$.
Jump increase in $\r_s d^2$ associated with layering shows up in sudden large
increase in conductance near $\mu=7.5$. Another sharp increase
 in $\r_s d^2$ near $\mu=12$
is due to growth of lattice planes by one atom each; this happens in the
orthogonal direction to heat conduction and does not affect $G_s$.
}
\label{lambda}
\end{figure}
In Fig.\ref{lambda} we show the heat conductance in solid region $G_s$ as
a function of strength of the trapping potential $\mu$. The inset in 
 Fig.\ref{lambda} shows the change in the averaged density of the solid 
region $\r_s d^2$.
The $\r_s d^2-\mu$ plot show clear staircase-like sharp increases near
$\mu=8,\, 12$. As $\mu$ is  increased from $7$ to $8$, a layering transition
in a direction perpendicular to the interfaces occurs; whereas as $\mu$ is 
increased from $10$ to $12$ each of the lattice planes lieing parallel to
the solid-liquid interfaces grow by one atom 
(see also Fig.\ref{dislo}(a) \& (c)). These two modes of density fluctuations
leave their signatures by enhancing heat conductance $G_s$. Notice that,
the layering transition at $\mu=8$ increases stress in the solid region in the 
direction of heat conduction, thereby showing a step-like increase. The
other mode of density fluctuation is in the normal direction to heat transport
and thereby affects the heat conductance only near the transition due to 
the associated enhancement of overall fluctuations (a sharp peak at $\mu=12$). 

We now find out the Kapitza
resistance across the solid liquid interface as a function of the
strength of the external potential $\mu$.
With increase in $\mu$, the system shows a
jump in the density of the solid region corresponding to the addition
of an entire layer of atoms (see inset of Fig.\ref{lambda}). 
From the profile shown in
Fig.~\ref{lte}(b), the Kapitza resistance is easily obtained by
dividing the temperature jump by the energy flux. Note that a slight
dependence of $R_K$ on ${T}$ is visible in Fig.~\ref{lte}(b) with
a larger temperature jump on the ``warm'' side. The results shown in 
Fig.\ref{kapr} correspond to the average values of $R_K$ over the ``warm'' 
and ``cold'' sides.
\begin{figure}[t]
\begin{center}
\includegraphics[width=7.cm]{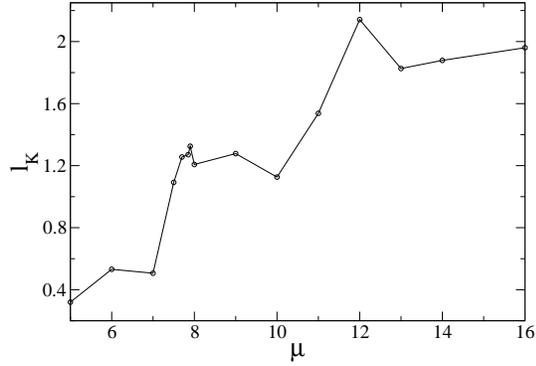}%{kapl-new.eps} %{Figures/spalitza-fig24.eps}
\end{center}
\caption{Plot of Kapitza length $l_K$ in units of $L_s$,  
as a function of $\mu$. This shows a jump increase at the layering transition.
}
\label{kapl}
\end{figure}
The plot of $R_K$ as a function of $\mu$ shows a distinct jump as a
layer is included (for a value of $\mu$ close to $8$) in the solid
region (Fig.~\ref{kapr}). The jump in $R_K$ is also
accompanied by a local dip at the transition, corresponding to larger
number fluctuations at the interface. 
The combined effect of the enhanced Kapitza resistance as
well as enhanced conductance of the solid can be summarized by defining 
the Kapitza length in units of width of the solid region $L_s$ as 
$l_K= R_K G_s$. This is a measure of excess width of a solid which is equivalent
in giving rise to a resistance equal to the Kapitza resistance. This is
reminiscent of the ``effective liquid" concept which has been used in
explaining some of the features of acoustic shock absorption in the
last section. The Kapitza length also shows a peak near the layering 
transition at $\mu=12$ (Fig.~\ref{kapl}).  

The layering transitions in solid occur via dislocation formation which
ultimately annihilates by incorporating more particles from the liquid
region. The kinematics of dislocation formation and annihilation is 
assisted by diffusion and dislocation climb which are very slow 
processes\cite{debc} in a solid compared to particle collision and 
kinetic energy transfer times. Thus it is possible for a system with metastable 
dislocation pairs to reach a thermal quasi-steady state. 
Fig.\ref{dislo}.(d) shows overlapped configurations of the solid region
containing a dislocation-antidislocation pair, as the system is  
quenched from $\mu=16$ to $\mu=24$. The overlapped configurations are
separated by time $100\,\d t$ and collected after a time of $15000\,\d t$
after the quench begins. Since dislocations annihilate though a conserved
diffusive dynamics which takes a long time ($10^5\d t$) compared to the
particle collision and kinetic energy transfer times, the system in presence
of dislocation pairs is in a effective steady state. It also maintains LTE 
that we have checked by computing $\la v^4(y)\ra$ and $\la v^2(y)\ra$ locally.
Thus, in a similar manner as stated before, we find
the conductance in solid region within this time scale when the solid is
decorated by the dislocation pair. This gives a heat conductance 
$G_s=2.29\, (\tau_s d)^{-1}$.
After a further wait for $10^5 \,\d t$ 
the dislocations get annihilated. 
At this stage the whole trapped region is transformed into
a $24$-layered solid. Then the heat conductance comes out to be 
$G_s= 3.53\, (\tau_s d)^{-1}$. 
Thus after complete annihilation of the
dislocation pair the conductance of the solid rises by about $54\%$!
This study already indicates towards the fact that dislocations behave like
large resistances towards heat transport. One can use the presence of 
stable dislocation-pairs as  have been obtained in confined narrow 
strips\cite{debc,debc-heat} for a more careful study of impact of dislocations
on heat transport.

\subsection{Approximate theory}
We now provide an approximate theoretical approach to calculate heat
conductance within the solid region. We use a free volume type calculation
to obtain an approximate estimate for heat conductance starting from
an exact expression for $\a$-th component of the heat flux 
density\cite{debc-heat} 
\bea
j_\a(\br) &=& j_\a^K(\br)+j_\a^U(\br)=\sum_i \d (\br - \br_i) h_i \bv_i^\a \nn\\
&+& \f{1}{2} \sum_{i,j \neq i} \theta (x_i^\a-x^\a)
\prod_{\nu  \neq \a} \d (x^\nu-x_i^\nu)f_{ij}^\be( v_i^\be + v_j^\be)\nn\\
\label{jr}
\eea
obtained  from continuity of local energy density. Here 
$\theta(x)$ is the Heaviside step function and $\d(\dots)$ is a
Dirac delta function, 
$h_i=m {\bf v}_i^2/2+\phi({\bf r}_i)+\sum_{i>j} u(r_{ij})$,
$\phi({\bf r}_i)$ is an onsite potential and $u(r_{ij})$ is 
the inter-particle interaction.
The first term in Eq.\ref{jr} $j_\a^K(\br)$ denotes convection while the 
second term $j_\a^U(\br)$ denotes conduction.
The above formula for conduction has an simple interpretation.
The sum is over only those $i$ for which $x_i^\a > x^\a$.
Thus this formula gives the net rate at
which work is done by particles on the left of $x^\a$ on the particles
on the right which is therefore the rate at which energy flows from left to
right.
The $\a$-th component of the integrated heat current density over the 
solid region\cite{debc-heat},
\bea
\la I_\a \ra = 
 \sum_i \la~ h_i v_i^\a \ra -\f{1}{4} \sum_{i ,j \neq
  i} \left\la~ \f{\p u(r_{ij})}{\p r_{ij}}
\f{x^\a_{ij} x^{\be}_{ij}}{r_{ij}} (v_i^\be + v_j^\be)~ \right\ra.\nn\\ 
\label{totI}
\eea

In solid region most of the transport is carried out by conduction and one
may ignore the convection part.
In this study we focus on the average heat current density along $y$-direction,
$j_E = \la I_y \ra/L_x L_s$.
Ignoring convection inside solid, approximating the system as a system
of hard disks with some effective hard disk diameter $\s$ and assuming 
LTE \cite{debc-heat}, the  heat conductance in units of $(\tau_s d)^{-1}$
can be expressed as,
\bea
G_s = \f{j_E}{\D T} = 
\left[3 \f{\r_s}{L_s} \f{y_c^2}{\tau_c}\right] \left(\f{d}{\s}\right)^2
\eea
where $\r_s=4\eta_s/\pi$, 
$y_c$ is the average separation between the colliding particles in 
$y$-direction, $\tau_c$ is the mean collision time and $\r_s$ is the 
average density of the solid. For the details of this derivation refer 
to Ref.\cite{debc-heat}.
The extra factor of $(d/\s)^2$ is due to the mapping of the soft disks
of diameter $d$ to effective hard disks of diameter $\s$.
  
\begin{figure}[t]
\begin{center}
\includegraphics[width=7.cm]{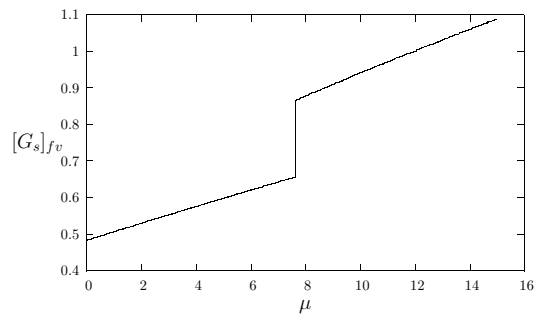}
\end{center}
\caption{Free volume estimate of the $y$- component of heat conductance 
$[G_s]_{fv}$, in units of $(\tau \s)^{-1}$, in the solid region of a 
hard disk system composed of liquid and trapped solid region. 
Heat conduactance shows jump increase as layering 
transition occurs with increase in trapping potential $\mu$.
}
\label{thcnd}
\end{figure}

Now this conductance $G_s$ can be calculated if one can obtain some
estimate for $y_c$ and $\tau_c$. We estimate $y_c^2$ and $\tau_c$
from the fixed neighbor free volume theory (FNFVT) as
in Ref.\cite{debc-heat}. For different values
of $(\eta_s,L_s)$ [~$\eta_s$ obtained by extremizing the free energy of
the full system at a given potential well depth $\mu$, 
as discussed in Sec.\ref{layer}~], one can obtain
$a_0,~\ve_{xx},~\ve_{yy}$ that gives the basic geometrical inputs of
$b$, and $h$ (Fig.\ref{fv}). We assume the test particle $P_0$ moves in the
cage formed by its neighbors and obtain the average values $[y_c^2]_{fv}$
and $[\tau_c]_{fv}$ for the moving particle from FNFVT. We assume that
the position of the centre of the moving disk $P_0(x,y)$, at the time
of collision with the other disks, is uniformly distributed on the 
boundary $\cal B$ of the free volume. Then $[y_c^2]_{fv}$ can be easily
calculated using the expression\cite{debc-heat}
\bea
[y_c^2]_{fv}=\f{\sum_i  \int_{{\cal{B}}_i} ds
  (y-y_i)^2}{L_{\cal{B}}}~,\label{ycfv} 
\eea 
where ${\cal{B}}_i$ is the part of the boundary $\cal{B}$ of the free
volume when the middle disk is in contact with the $i^{\rm{th}}$ fixed
disk, $ds$ is the infinitesimal length element on $\cal{B}$ while
$L_{\cal{B}}$ is the total length of $\cal{B}$.
An exact calculation of $[\tau_c]_{fv}$ is
nontrivial.  
However we expect $[\tau_c]_{fv}=c~v_{fv}^{1/2}/T^{1/2} $ where
$v_{fv}$ is the ``free 
volume'' [see Fig.~\ref{fv}] and   $c$ is a constant factor of
$O(1)$ which may be used as a fitting parameter. Thus the heat 
conductance in solid region may be expressed as,
\bea
[G_s]_{fv} = \f{ 3 \r_s \s^2 T^{1/2}}{L_s d^2} ~\f{[y_c^2]_{fv}}{c ~v_{fv}^{1/2}}\,. 
\eea
There are well defined schemes\cite{hansen} to approximate a soft disk
system as a system of effective hard disks. However, we choose a
much simpler path of finding the heat conductance of a trapped hard disk
system instead. In Fig.\ref{thcnd}
we plot $[G_s]_{fv}$ as  a function of $\mu$ with $L_s=19.73\s$
the same width of the trapped hard disks used in Sec.\ref{hdint}.
For hard disks we obtain heat conductance in units of $(\tau\s)^{-1}$ and
use an average temperature $\kb T=1$ and set $c=1$.
This gives the estimate of heat conductance along $y$-direction in a 
hard disk system that showed a layering transition from $21$ to $22$ layers
near $\mu=8$ in Fig.\ref{step}.
The plot clearly shows the associated increase in heat conductance in
the solid region. This behavior is also in qualitative agreement with 
Fig.\ref{lambda}.
 
With the help of these results we may conclude that the layering
transition has a profound effect on the thermal properties of the
trapped solid lieing in contact with its liquid. 
An important consequence of this study is the
possibility that the thermal resistance of interfaces may be altered
using external potential which cause layering transitions in a trapped
nano solid. 
Moreover, the heat conduactance of the solid may be drastically reduced
by tuning the trapping potential. 
We believe that these phenomena have the potential for
useful applications for e.g. as tunable thermal switches or in other nano
engineered devices. 

\section{Conclusion}

In this paper we have shown that liquid-solid interfaces which are 
constrained by strong chemical potential gradients remain flat and at 
the same time undergo fluctuations which increases or decreses the number of 
solid layers by one. The nature of these fluctuations are strongly influenced 
by the size of the solid. It is expected that for macroscopically large solids 
these fluctuations would reduce to the random nucleation of steps on the 
solid surface and the sort of coherence observed here would be absent. 
Confining a thin `long' strip of solid by smooth walls in quasi one dimension 
leads, strictly speaking, to a 
destruction of solid-like order\cite{andrea,debc} and an enhancement of 
smectic-like ordering of individual layers parallel to the confining walls. 
It is this reduction of interlayer coupling which is ultimately responsible 
for the spallation of single solid layers. 

Before we end, we would like to examine carefully the relevance of our 
results to practical situations and experimental systems constructed in 
a laboratory. 
In recent years our ability to manipulate matter at an atomic or molecular 
level has increased tremendously\cite{nano}. It is possible now to 
localize atoms using 
carefully designed atomic traps\cite{trap,phillips} and observe their 
properties.
It is also 
possible to set up experiments where atoms may be picked one by one and 
arranged in any specified pattern\cite{nano}. Parallel to this development, 
one can 
now synthesize functionalized colloidal\cite{hamley,alfons} particles with 
a variety of shapes 
and sizes which mimic the properties of atomic matter at length and time 
scales which are easy to handle even in relatively inexpensive experimental 
set-ups. Colloidal particles can also be manipulated using laser tweezers 
and traps\cite{baumgartl,phillips,collayer,col-trap}, confining within narrow 
channels and slits\cite{collayer,col-conf}, adsorbing on substrates 
or air-water interfaces\cite{Zahn} and assembling layer by layer 
using carefully designed substrate templates\cite{AMOLF,col-subs}. 

The structural aspects of our results viz. the layering 
transition\cite{aypa,collayer} and 
all associated phenomena should be observable in both atomic systems 
like rare gases in atom traps and in colloids on templates or in laser 
fields. Indeed, 
interfacial fluctuations similar to the sort discussed in this 
paper have been observed during early nano-indentation 
experiments\cite{nindent}. Layering transitions have also been observed in 
shaken hard disks more than two decades ago\cite{pieransky}. In confined 
molecular systems such layering transitions are of great relevance to 
the study of nano-tribology\cite{aypa}. 

The dynamical aspects of our results, on the other hand, will be difficult to observe in 
colloidal systems because of viscous damping by the solvent. Nevertheless,
it may still be possible to observe some of these effects if this damping is 
small\cite{shriram}. Our results for momentum and heat conduction in 
trapped solids therefore pertain mainly to atomic or molecular systems where
such damping is absent\cite{aypa}. 
As such, we do not see any difficulty for generalizing the main conclusions of 
our study to dimensions higher than two. In three dimensions, confinement to a 
slit should have similar effects on a three dimensional solid viz. reducing 
interlyer coupling so that two dimensional layers may become partially 
independent. Weak shock waves may then be able to spallate these layers into
a surrounding liquid or gaseous phase. Work along these lines is in progress.

\acknowledgements
We would like to thank Madan Rao, Abhishek Dhar, Tamoghna K. Das, 
Kurt Binder, Andrea Ricci and Peter Nielaba for discussions
and Martin Zapotocky for a critical reading of the manuscript. 
This work was partially supported by 
the Department of Science and Technology and CSIR (India).

\end{document}